\documentclass{emulateapj}
\usepackage{graphics,graphicx}
\usepackage{subfigure}
\usepackage{epsfig}
\usepackage{natbib}
\usepackage{amssymb}
\usepackage{lscape}

\shorttitle{The ICMART Model of GRBs} 
\shortauthors{Zhang \& Yan}
\begin{document}
\title{The Internal-Collision-Induced Magnetic Reconnection
and Turbulence (ICMART) Model of Gamma-Ray Bursts}
\author
{Bing Zhang\altaffilmark{1}, Huirong Yan\altaffilmark{2,3}}
\altaffiltext{1}{Department of Physics and Astronomy, University
of Nevada Las Vegas, Las Vegas, NV 89154.}
\altaffiltext{2}{Kavli Institute of Astronomy and Astrophysics,
Peking University, Beijing 100871, China.}
\altaffiltext{3}{TAP Fellow, University of Arizona, 1629 E.
University Blvd., Tucson, AZ, 85721.}

\begin{abstract}
The recent Fermi observation of GRB 080916C shows that the 
bright photosphere emission associated with a putative fireball
is missing, which suggests that the central engine likely launches 
a Poynting-flux-dominated outflow. 
We propose a model of gamma-ray burst (GRB) prompt emission
in the Poynting-flux-dominated regime, namely,
the Internal-Collision-induced MAgnetic Reconnection and 
Turbulence (ICMART) model. It is envisaged that the GRB central 
engine launches an intermittent, magnetically-dominated wind,
and that in the GRB emission region, the ejecta is still 
moderately magnetized (e.g. $1\lesssim \sigma \lesssim 100$). 
Similar to the internal shock (IS) model, the
mini-shells interact internally at the radius 
$R_{\rm IS}\sim \Gamma^2 c \Delta t$. 
Most of these early collisions, however, have little energy 
dissipation, but serve to distort the ordered magnetic field
lines entrained in the ejecta.
At a certain point, the distortion of magnetic field 
configuration reaches the critical condition to allow fast 
reconnection seeds to occur, which induce relativistic MHD 
turbulence in the interaction regions. The turbulence further 
distorts field lines easing additional magnetic reconnections,
resulting in a runway release of the stored magnetic field energy 
(an ICMART event). 
Particles are accelerated either directly in the reconnection zone, 
or stochastically in the turbulent regions, which radiate 
synchrotron photons that power the observed gamma-rays. 
Each ICMART event corresponds to a broad pulse in the GRB light curve,
and a GRB is composed of multiple ICMART events.
This model retains the merits of the IS and other models, but 
may overcome several difficulties/issues faced by the IS 
model (e.g. low efficiency, fast cooling, electron number excess,
Amati/Yonetoku relation inconsistency, and missing bright photosphere).
Within this model, the observed GRB variability time
scales could have two components, one slow component 
associated with the central engine time history, and another
fast component associated with relativistic magnetic turbulence
in the emission region. The model predicts a decrease of gamma-ray 
polarization degree and $E_p$
in each ICMART event (broad pulse) during the prompt GRB phase,
as well as a moderately magnetized external reverse shock. The 
model may be applied to the GRBs that have time-resolved, 
featureless Band-function spectra, such as GRB 080916C and most 
GRBs detected by Fermi  LAT.
\end{abstract}


\section{Introduction}

The composition of the Gamma-Ray Burst (GRB) ejecta has remained
a mystery until recently. The uncertainty lies in the lack of
knowledge about the ratio between Poynting flux and matter 
(baryonic) flux\footnote{Strictly speaking, the matter flux is 
the sum of the baryonic flux and the leptonic flux. It is usually
dominated by the baryonic flux unless the pair number density is 
so high that $N_{\pm}/N_b \geq m_p/m_e$ is satisfied.},
i.e.\footnote{The last equation applies to the
case that the magnetic field lines are perpendicular to the 
direction of the motion, which is the configuration invoked in
the ICMART model proposed in this paper. If magnetic fields are 
generated in the internal shocks due to plasma instabilities, 
the comoving field lines would have random orientations. In 
this case, one has $\sigma \simeq {B'}^2/8\pi \rho' c^2$.}
\begin{equation}
\sigma = \frac{F_{P}}{F_{b}} = \frac{B^2}{4\pi\Gamma\rho c^2}
=\frac{{B'}^2}{4\pi{\rho}' c^2}~,
\label{sigma}
\end{equation}
where $B$ and $\rho$ are the magnetic field strength
and matter density in the rest frame of the central engine
(or the lab frame), and $B'$ and $\rho'$
are the corresponding quantities in the rest frame comoving with 
the ejecta. The standard picture is the ``fireball'' shock 
model \citep{paczynski86,goodman86,shemi90,rees92,meszarosrees93,rees94}. 
Within such a picture, an initially hot fireball composed of
photons, electron-positron pairs, and a small amount of baryons
first converts most of its thermal energy into kinetic energy,
and then dissipates the kinetic energy in the internal (or sometimes
external) shocks to power the observed GRB emission. Within 
such a scenario, the magnetic field is assumed not to play a 
dynamically important role in the ejecta, i.e. $\sigma \ll 1$. 
Such a field can be generated 
{\em in-situ} via plasma instabilities 
\citep{weibel59,medvedev99,nishikawa05,nishikawa09,spitkovsky08}
in relativistic shocks,
in which Fermi-accelerated electrons cool via synchrotron/jitter
\citep{meszaros94,tavani96,medvedev00} or synchrotron self-Compton
(SSC) \citep{meszaros94,kumarmcmahon08} radiation.
An alternative view is that the GRB ejecta carries a dynamically
important magnetic field component, i.e. $\sigma \gg 1$.
The GRB radiation is powered by dissipation of the 
magnetic field energy in the ejecta 
\citep{usov92,thompson94,meszarosrees97b,lyutikov03,vlahakis03,lazarian03,lyutikov06}.

Until recently, it has been difficult to diagnose the composition
of GRB ejecta. Regardless of the $\sigma$ values, the late time 
afterglow behavior appears the same, which is the emission of 
the ``forwardly'' shocked circumburst medium. Successful 
modeling of some afterglow data therefore does {\em not} shed 
light onto the composition of the ejecta. An important phase is 
when the ejecta energy is transferred to the circumburst medium. 
This is usually accompanied by passing of a reverse shock across 
the ejecta \citep{sari95,zhangkobayashi05}. The brightness of the 
reverse shock emission is found to be dependent on the magnetization 
parameter \citep{fan04b,zhangkobayashi05,mizuno09,mimica09}.
However, both low-$\sigma$ and high-$\sigma$ flows
can lead to a relatively dim reverse shock emission
\citep{zhangkobayashi05,jinfan07}, so that it is
difficult to robustly constrain $\sigma$ from the 
observations of the reverse shock emission. Finally, the prompt 
emission light curves and the narrow-band spectra (in the energy 
bands of the previous gamma-ray detectors such as BATSE, Swift, etc)
may be accommodated within the frameworks of both the low-$\sigma$ 
(internal shock) and high-$\sigma$ (magnetic dissipation) 
models, so that they do not carry adequate information
to diagnose GRB composition. The progress in constraining GRB 
composition therefore has been slow.

In any case, it has been long speculated that the GRB ejecta
are somewhat magnetized.  The putative
GRB central engines, either a black hole - torus system 
\citep[e.g.][]{proga03} or a rapidly spinning neutron star 
\citep[e.g.][]{usov92}, are very likely magnetized. 
The GRB ejecta is likely entrained
with a globally structured magnetic field. In principle, the 
$\sigma$ factor can reach $\sim 1$ or even $\gg 1$.
Before the launch of Fermi, several authors have argued
for a strongly magnetized GRB central engine based on some
observational evidence and its theoretical modeling.
\cite{coburn03} reported a $80\% \pm 20\%$ degree of linear
polarization in GRB 021206, which is best interpreted by 
synchrotron emission in a globally structured magnetic field
\citep{waxman03,lyutikov03b,granot03}. However, a further 
analysis of the same data cannot confirm the claim
\citep{rutledge04}. By modeling early optical flashes within
the reverse shock model, \cite{zhang03} found that the
early optical flashes with a rapidly decaying light curve (reverse 
shock component) followed by a flattening feature (forward
shock) as observed in GRB 990123 and GRB 021211 generally
require that the reverse shock region is much more magnetized
than the forward shock region. Such a conclusion was 
independently drawn by \cite{fan02} and \cite{kumar03}
through detailed case studies of the two GRBs. This extra
magnetization has to be related to a magnetized central engine.
More recently, \cite{kumarmcmahon08,kumarpanaitescu08} analyzed
the prompt emission data of several GRBs based on a largely 
model-independent method. They concluded that the observed
emission cannot be produced by synchrotron emission within
the internal shock model. They tentatively suggested a
magnetized ejecta as the source of the observed gamma-ray
radiation.

Recent Fermi observations shed light onto the composition of
some GRBs. Fermi carries the Gamma-ray Burst
Monitor (GBM) and the Large Area Telescope (LAT), which
cover a broad spectral range of 6-7 orders of magnitude in energy.
It is ideal to measure the broad-band spectra of GRBs during
the prompt emission phase. The first bright GRB co-detected
by GBM and LAT, GRB 080916C, showed several nearly featureless,
smoothly-joint-broken-power-law spectra (the so-called 
``Band''-function, \cite{band93}) covering
6-7 decades in energy \citep{abdo09,zhang10}. Although this might
not be surprising for observers (the Band-function has been
known since the early BATSE years), it was somewhat surprising
for modelers, since 
according to the standard fireball model, a thermal component 
associated with the fireball ``photosphere'' is expected
to be very bright and should be detected \citep{zhangpeer09}.
This component is analogous to the cosmic microwave background
radiation (CMB) associated with the hot Big Bang, and is
predicted to be bright enough to be detectable within a wide range of
parameters \citep{meszarosrees00,meszaros02b,peer06,peer08}.
\cite{zhangpeer09} analyzed this burst and argued
that the non-detection of the thermal component strongly suggests 
that the majority of energy (more than $95\%$) was not stored
in the form of a ``fireball'' at the central engine, but was 
stored in magnetic fields which was not released until reaching
a large radius. This
suggests that at least for GRB 080916C the outflow has to be 
{\em Poynting flux dominated} (PFD) with $\sigma > (15-20)$ 
at the central engine and at the photosphere as 
well. A follow up investigation \citep{fan10} on the various
possibilities of hiding the thermal component (e.g. by invoking 
a smaller central engine radius, which is not required by 
the minimum variability time scale data of GRB 080916C) 
confirms the conclusion \citep[see also][]{gao09}\footnote{
An alternative model is to interpret the entire spectra
as the emission from a dissipative fireball photosphere
\citep{beloborodov09,lazzati10}. This model, although plausible
to interpret some GRBs, is found to be difficult to interpret the 
data of GRB 080916C, see \citep{zhang10} and Sect.2.3 for a more 
detailed discussion).}. All these call for 
a serious re-investigation of GRB prompt emission models
in the high-$\sigma$ regime. 

In this case, magnetic energy may be sufficient to feed GRBs. 
Magnetic reconnection was suggested as a component for GRBs 
long ago (Thompson 1994). The problem lied, however, in the 
intrinsic difficulty of reconnection as it is a very slow process 
in ordered fields. As with the case for solar flares, both a slow 
phase of accumulation of the oppositely directed flux and a fast 
bursty phase are required for reconnection. Essential progress 
was made by \citep{lazarian03}, who proposed a new scenario for GRBs 
by invoking  
self-adjusted reconnection based on the findings of fast reconnection 
in 3D turbulent magnetic fields \citep{lazarian99}. They suggested 
that the fast bursty reconnection eventually occurs as a nonlinear 
feedback of the increased stochasticity of the magnetic field lines. 
The reconnection events start from some limited volumes and then 
spread in the form of a chain reaction as the energy is fed back 
to the turbulence and induces dramatic change in the magnetic field 
topology. The turbulent reconnection model has been confirmed by 
recent numerical testings (Kowal et al. 2009). 

In this paper, we build a model of GRB prompt emission
in the high-$\sigma$ regime on the basis of the recent observational 
and theoretical achievements. This model is called the
{\em Internal-Collision-induced MAgnetic Reconnection and
Turbulence} (ICMART) model. In \S\ref{sec:models}, we critically
review the current prompt emission models, including the 
internal shock (IS) model (\S\ref{sec:IS}) in the low-$\sigma$
regime, the electro-magnetic model in the 
extremely high-$\sigma$ regime (\S\ref{sec:EM}) and the 
dissipative photosphere model in the low- to intermediate-$\sigma$ 
regime (\S\ref{sec:MHD}). We argue that these models are not
ideal to interpret GRB 080916C and some other GRBs.
We then delineate the general picture of the
ICMART model in \S\ref{sec:ICMART}. In \S\ref{sec:merits} 
we discuss the merits of the ICMART model, 
focusing on how it inherits the merits of the previous
models and overcomes their drawbacks. 
In \S\ref{sec:predictions}, we outline several unique 
predictions of the ICMART model, which can be used to 
differentiate this model from other models based on the
observational data. These include the existence of
two variability components, an evolution of $E_p$ and 
gamma-ray polarization degree across each GRB pulse, as
well as a mildly polarized external reverse shock emission 
from GRBs. We then dedicate \S\ref{sec:GRB080916C} discussing
how the ICMART model interprets the observation of GRB 080916C.
The ICMART model is summarized in \S\ref{sec:conclusion} with 
some discussion on its broad implications on other aspects of
GRB physics (e.g. neutrino and cosmic ray connections with GRBs)
as well as its possible applications to other astrophysical
objects such as active galactic nuclei (AGNs). 
The physics invoked in this model (e.g. turbulence and 
reconnection in a high-$\sigma$, relativistic plasma) is complicated.
In this paper we make the first step by delineating the qualitative 
picture of the model, and defer the more detailed quantitative/numerical 
analyses to future work. 

\section{Previous GRB prompt emission models}\label{sec:models}

\subsection{The internal shock model}\label{sec:IS}

For a baryonic outflow ($\sigma \ll 1$), the standard GRB 
prompt emission model is the internal shock (IS) model
\citep{rees94,paczynski94}. It is envisioned that the
GRB central engine launches an unsteady
outflow with varying luminosity and
Lorentz factors. Approximating the outflow as
a series of ``mini-shells'' with a distribution of Lorentz 
factors [$\Gamma \in(\Gamma_{\rm min}, \Gamma_{\rm max})$],
widths [$\Delta \in(\Delta_{\rm min}, \Delta_{\rm max})
=(c \Delta t_{\rm min}, c \Delta t_{\rm max})$, where
$\Delta t$ is the duration of the central engine activity of
each minishell], and separations [$d \in(d_{\rm min}, d_{\rm max})
=(c \delta t_{\rm min}, c \delta t_{\rm max})$, where
$\delta t$ is the duration between the end of ejecting
a leading shell and the beginning of ejecting a trailing
shell], one can obtain a series of collisions due
to the interactions between the late, fast shells and the
early, slow shells. These collisions are supersonic, resulting
in internal shocks from which particles are accelerated 
and photons are released to power the GRB prompt emission.
For two shells with parameters 
$(\Gamma_s, \Delta_s)$ and $(\Gamma_f, \Delta_f)$ separated
by $d = c \delta t$ (with the fast shell lagging behind the
slow shell), the internal shock radius is (noticing 
$\beta=(1-\Gamma^{-2})^{1/2}$)
\begin{equation}
R_{\rm IS} = \frac{d}{\beta_f - \beta_s} \simeq 
2 \Gamma_s^2 c \delta t
= 6\times 10^{14}~{\rm cm}~ \Gamma_{s,2.5}^2 \delta t_{-1}~,
\label{RIS}
\end{equation}
where $\Gamma_{s,2.5} = \Gamma_s/10^{2.5}$. Hereafter the 
convention $Q_s = Q/10^s$ is adopted in cgs units 
throughout the text.

\subsubsection{Merits: central-engine-driven variability}
Most GRB light curves are highly variable. The internal shock
model attributes this variability to that of the central engine.
Internal collisions are also frequently observed 
or inferred from other astronomical objects, such as pulsar wind 
nebula, AGN, and planetary nebulae. It is therefore natural
to envision internal collisions in GRBs as well.
The connection between the 
observed GRB variability and that of the GRB central engine is 
recently strengthened by the
observations and modeling of X-ray flares found in some
Swift GRBs \citep{burrows05,falcone06,romano06,chincarini07,falcone07},
which are believed to be due to late central engine activities
\citep{burrows05,zhang06,fanwei05,lazzati07,maxham09}. A strong 
support to such an interpretation was presented by \cite{liang06}, 
who blindly searched for $T_0$ of the flares 
based on the high-latitude ``curvature'' effect model of the 
decaying phase of the flares using the observed temporal
and spectral data. They found that the required $T_0$ are 
often near the beginning of the flares, which strongly
suggests that the GRB central engine is restarted. Since X-ray flares
and gamma-ray pulses share the same origin, as is demonstrated by
the smooth transition between the two phases \citep[e.g.][]{krimm07},
the GRB/X-ray flare data in general demand that the observed 
variability in GRBs should be tied to that of the central engine.
The internal shock model naturally makes such a connection, i.e. 
the observed variability time history roughly traces the time 
history of the central engine \citep{kobayashi97,maxham09}. 
Furthermore, it has been argued that the radiation efficiency 
of the internal shock model can be much higher than that of the 
external shock model to interpret variable light curves 
\citep{sari97} cf. \citep{dermermitman99} (but still not
efficient enough to interpret the data, see below). 
This has made the IS model a popular theoretical model for
GRB prompt emission for many years.

Despite of its popularity, the IS model suffers from several
criticisms, which we summarize in the following.

\subsubsection{The efficiency problem}

Suppose that the second, fast shell ($m_2, \Gamma_2$) catches up 
with the first, slow shell ($m_1, \Gamma_1$), and that the two 
shells undergo a 
full inelastic collision and generate an internal energy $U'$.
From energy conservation, $\Gamma_1 m_1 + \Gamma_2 m_2
=\Gamma_m (m_1+m_2+U'/c^2)$, and momentum conservation,
$\Gamma_1\beta_1 m_1+\Gamma_2\beta_2 m_2 =\Gamma_m
\beta_m (m_1+m_2+U'/c^2)$, one can derive the Lorentz
factor of the merged shell
\begin{equation}
\Gamma_m = \left( \frac{\Gamma_1 m_1 + \Gamma_2 m_2}
{m_1/\Gamma_1 + m_2/\Gamma_2} \right)^{1/2}~.
\label{Gammam}
\end{equation}
The energy dissipation efficiency (which is the upper
limit of the radiation efficiency when the ``fast cooling''
condition is satisfied) of the collision is
\begin{eqnarray}
\eta_{_{\rm IS}} & = & \frac{\Gamma_m U'}{\Gamma_1 m_1c^2+\Gamma_2 m_2c^2}
\nonumber \\
& = & 1 - \frac{m_1+m_2}{\sqrt{m_1^2+m_2^2 + m_1m_2
\left(\frac{\Gamma_2}{\Gamma_1}+\frac{\Gamma_1}{\Gamma_2}
\right)}}~.
\label{etaIS}
\end{eqnarray}
This efficiency is typically low 
\citep{kobayashi07,panaitescu99,kumar99}. 
The most efficient collisions are those with equal masses
and large $\Gamma$ ratios, but for a distribution of 
mass and Lorentz factor of a group of randomly ejected 
shells, the mean efficiency is low, e.g. $<15\%$ for
$\Gamma_{\rm max} / \Gamma_{\rm min} < 100$. For a standard
electron equipartition parameter $\epsilon_e \sim 0.1$, this
model predicts an X-ray afterglow flux brighter than the
observed flux by near 3 orders of magnitude \citep{maxham09}.
Increasing $\eta_{_{\rm IS}}$ requires adjusting the
Lorentz factor distribution of the shells, so that the
$\Gamma$ contrast is systematically increased 
\citep{beloborodov00,kobayashisari01,guetta01}. There is no
obvious physical reason why GRB central engines would 
satisfy these contrived requirements. Observationally,
a detailed study of the GRB radiative efficiency based on
the Swift BAT/XRT data of a sample of GRBs \citep{zhang07a} 
suggests that the radiative efficiency of some GRBs can be 
as high as $90\%$. This is difficult to interpret within
the straightforward IS models. 

\subsubsection{The fast cooling problem}
\label{sec:fastcooling}

In the internal shock scenario, once an electron is accelerated
at the shock front, it would cool rapidly downstream
through synchrotron/SSC emission.
A typical GRB has an observed peak energy $E_p \sim 250$ keV. 
This can be translated into an estimate of the typical electron 
Lorentz factor $\gamma_{e,p}$ that contributes to $E_p$. Within 
the the standard synchrotron emission model, one has 
\begin{equation}
E_p \sim \hbar \Gamma \gamma_{e,p}^2 \frac{e B'}{m_e c} (1+z)^{-1}~,
\label{Epsyn}
\end{equation}
where $\Gamma$ is the bulk Lorentz factor of the shocked region
($\Gamma_m$ in Eq.[\ref{Gammam}] for an individual collision in
the IS model); $B'$ is the magnetic field strength in the comoving 
frame. For an outflow (or ``wind'') with a mean kinetic luminosity 
$L_w$, the total internal energy due to shock dissipation is $L_w 
\eta_{_{\rm IS}}$. Assuming that this internal energy is
distributed to protons, electrons and magnetic fields in
the fractions $\epsilon_p$, $\epsilon_e$ and $\epsilon_B$,
with $\epsilon_p + \epsilon_e + \epsilon_B = 1$, one has
$(L_w \eta_{_{\rm IS}} \epsilon_B)/(4\pi R^2 c \Gamma^2) =
{B'}^2/8\pi$, so that $\Gamma B' = (2 L_w \eta_{_{\rm IS}}\epsilon_B
/c R^2)^{1/2}$. As will be evidenced soon below, the emission
is in the ``fast cooling'' regime so that the energy distributed
to electrons is essentially converted to the observed gamma-rays.
The (isotropic) gamma-ray luminosity (a direct observable) can
be approximated as $L_\gamma=L_w \eta_{_{\rm IS}} \epsilon_e$.
Taking a typical redshift $z=1$, Eq.(\ref{Epsyn}) gives the 
constraint
\begin{eqnarray}
\gamma_{e,p} & \simeq & 2.3 \times 10^3 L_{\gamma,52}^{-1/4} R_{14}^{1/2}
\nonumber \\
& \times & \left(\frac{\epsilon_e}{\epsilon_B}\right)^{1/4} 
\left(\frac{1+z}{2}\right)^{1/2} 
\left(\frac{E_p}{250~{\rm keV}}\right)^{1/2}~.
\label{gammaep}
\end{eqnarray}
The comoving cooling time scale for an electron with Lorentz
factor $\gamma_e$ is $t'_c (\gamma_{e,p}) = (\gamma_e m_e c^2)$ $/[(4/3)
\gamma_e^2 \sigma_T c ({B'}^2/8\pi)(1+{\cal Y})] =(6\pi m_e c)/[\gamma_e
\sigma_T {B'}^2(1+{\cal Y})] = 0.008~{\rm s}~ \gamma_{e,3}^{-1} {B'_4}^{-2}
(1+{\cal Y})^{-1}$, where ${\cal Y}=U'_{ph}/U'_{B}$ is  
the ratio between the comoving photon energy density and 
the magnetic density. Comparing with the comoving
dynamical time scale $t'_{dyn} = R/\Gamma c \sim 110~{\rm s}~
R_{15} \Gamma_{2.5}^{-1}$, one has $t'_c (\gamma_{e,p}) \ll t'_{dyn}$. 
This suggests that electrons at $\gamma_{e,p}$ cool
rapidly within the dynamical time, forming an electron spectrum
of $N(\gamma_e) \propto \gamma_e^{-2}$ below $\gamma_{e,p}$ all
the way to $\gamma_{e,c} \sim 1 \ll \gamma_{e,p}$ for the 
nominal parameters adopted. One therefore expects that
the photon number density below $E_p$ follows $N(E) \propto 
E^{-3/2}$ \citep{sari98}. This is different from the typical 
low energy photon spectrum observed in most GRBs, $N(E)
\propto E^{\alpha}$, with $\alpha \sim -1$ \citep{preece00}.
This is the ``fast cooling problem'' of the IS model 
\citep{ghisellini00}. A similar argument from a different
approach was presented in \cite{kumarmcmahon08}.
Since it stems from the synchrotron
cooling argument in general, this problem applies to all the
scenarios in which electrons are accelerated only once without 
continuous heating, such as the shock acceleration scenario 
commonly discussed in the literature. Possibilities to
alleviate this problem within the IS model
include introducing decay of the shock generated magnetic 
fields \citep{peerzhang06} and introducing 2nd order stochastic
Fermi acceleration in the post shock region due to plasma 
turbulence \citep{asano09}.

\subsubsection{The electron number excess problem} 
\label{sec:enumber}

In the internal shock model, if all the electrons associated
with the ejecta are accelerated, for typical parameters one has
too many electrons to share the dissipated internal energy,
so that the typical synchrotron emission frequency is smaller 
than what is observed by $\sim$ 2 orders of magnitude. The 
argument is the following.

Since the synchrotron spectrum for the internal shock
scenario is in the fast cooling regime, $E_p$ should be defined 
by synchrotron emission of the
minimum energy electrons injected at the shock front, 
i.e. $\gamma_{e,p} = \gamma_{e,m}$. 
The mean proton Lorentz factor $\bar\gamma_p$ depends on 
the parameters $(m, \Gamma)$ of the two colliding shells,
and the relative Lorentz factor between the two 
shells $\Gamma_{fs}=(\Gamma_f/\Gamma_s+\Gamma_s/\Gamma_f)/2$. 
Rigorously, one can divide the shock interaction
region into 4 regions (1: unshocked leading shell; 2: shocked
leading shell; 3: shocked trailing shell; 4. unshocked trailing
shell). The reverse shock is typically stronger than the 
forward shock, which is more relevant to GRB prompt emission. 
For a rough estimate, one has
\begin{equation}
\bar\gamma_p-1 \simeq (\Gamma_{43}-1) \epsilon_p~,
\label{bargammap}
\end{equation}
where $\Gamma_{43}$ is related to $\Gamma_{fs}$ and the
parameters  $(m, \Gamma)$ of the two colliding shells.
This can be reduced to $\bar\gamma_p \simeq \Gamma_{43}$ for 
$\epsilon_p \sim 1$. With the definitions of $\epsilon_e$ and 
$\epsilon_p$, one has $\bar \gamma_e-1 = (\epsilon_e n_p/\epsilon_p n_1) 
(m_p/m_e)(\bar \gamma_p-1)$.
The minimum Lorentz factors of electrons and protons can be written
as $(\gamma_{e,m}-1) = \phi(p) (\bar\gamma_e-1)$, and 
$(\gamma_{p,m}-1) = \phi(p)(\bar\gamma_p-1)$, respectively, 
with $\phi(p) \simeq (p-2)/(p-1)$ for $p>2$, and $\phi(p) = 
[\ln(\gamma_{M}/\gamma_m)]^{-1}$ for $p=2$ ($\gamma_M$
and $\gamma_m$ are the maximum and minimum Lorentz factors of the
power law distribution of the protons or electrons, respectively). 
Noticing $\hbar e B_q/m_e c = m_e c^2$, where $B_q =4.414\times
10^{13}~{\rm G}$, one can finally express Eq.(\ref{Epsyn}) into
\begin{eqnarray}
E_{\rm p,IS} & \sim & \frac{m_e c^2}{B_q} [\phi(p)]^2
\left(\frac{m_p}{m_e}\right)^2 
\left(\frac{2 L_\gamma}{R_{\rm IS}^2 c}\right)^{1/2} (1+z)^{-1}
\nonumber \\
& & \times \left[
\left(\frac{\epsilon_B}{\epsilon_e}\right)^{1/2}
\left(\frac{\epsilon_e n_p}{\epsilon_p n_e}\right)^2 (\bar\gamma_p-1)^2
\right] 
\nonumber \\
& \simeq & 4.4~{\rm keV}~ \left[\frac{\phi(p)}{1/6}\right]^2
L_{\gamma,52}^{1/2} R_{\rm IS,14}^{-1} \left(\frac{1+z}{2}\right)^{-1}
\nonumber \\
& & \times \left[
\left(\frac{\epsilon_B}{\epsilon_e}\right)^{1/2}
\left(\frac{\epsilon_e n_p}{\epsilon_p n_e}\right)^2 (\bar\gamma_p-1)^2~.
\right]~.
\label{Epsyn-IS}
\end{eqnarray}
Since in internal shocks one expects $\epsilon_B \sim \epsilon_e^2 < 
\epsilon_e$ \citep{medvedev06}, $\epsilon_e \ll \epsilon_p$, and
$(\bar\gamma_p-1) \sim 1$ (Eq.[\ref{bargammap}], unless a very large 
$\Gamma_{fs}$, and hence, $\Gamma_{43}$ is invoked), the typical 
$E_p$ predicted in the standard IS synchrotron emission model is 
about 2 orders of magnitude lower than the observed value. In order 
to invoke synchrotron radiation as the mechanism to power GRB prompt
emission, one needs to make an additional assumption
$n_p/n_e \gg 1$, i.e. only a small fraction of electrons are 
accelerated \citep{bykov96,daigne98}.

The requirement of a small fraction of accelerated electrons is
also introduced in correctly calculating the synchrotron 
self-absorption frequency ($\nu_a$) in the IS model. \cite{shen09} 
(see their Appendix for details) showed that in order to get a 
self-consistent $\nu_a$ using two
independent methods (one derived from the blackbody approximation
in the self-absorbed regime and another derived from the standard
approach of applying electron energy distribution \citep{rybicki79},
one generally requires that not all electrons associated with the
baryonic ejecta are accelerated. 

By modeling particle acceleration in relativistic shocks using
particle-in-cell (PIC) simulations, 
\cite{spitkovsky08} indeed showed that only a small fraction of
electrons are accelerated, and that most electrons ($\sim 99\%$
in number and $\sim 90\%$ in energy) are distributed in a
relativistic Maxwellian. The numerical simulation shows 
growth of the non-thermal tail with simulation time. 
\cite{spitkovsky08} suspects that the thermal peak may eventually 
get ``significantly eroded''. In any case, it would be interesting
to investigate the observational evidence of the putative thermal 
bump. The prompt GRB spectrum is usually well
fit by a Band function \citep[e.g.][]{abdo09}, which does 
not show a thermal-like bump near $E_p$ (which is believed
to be related to the injection energy of electrons due to the
fast cooling argument presented in Sect.\ref{sec:fastcooling}).
This suggests that the putative thermal bump may not be
significant (if it exists at all)
in the internal shocks within the IS model. \cite{giannios09}
interpret the early X-ray afterglow steep decay phase with
significant spectral evolution \citep{zhangbb07} as due to 
sweeping of the thermal bump across the X-ray band. This
interpretation inevitably attributes the emission phase before
the steep decay phase also to the external shock origin.
Observationally it is established that the X-ray steep decay
phase is a natural extension of the prompt emission 
\citep{tagliaferri05,barthelmy05b,obrien06}. The erratic
temporal behavior of the prompt emission is difficult to be
accounted for within the same external shock model that 
interprets the X-ray afterglow. A more natural interpretation
would be that the steep decay is the high-latitude emission
of the prompt emission as long as the instantaneous spectrum
at the end of the prompt emission phase is characterized by
a ``curved'' spectrum instead of a simple power law
\citep{zhangbb09}.

\subsubsection{The Amati/Yonetoku relation inconsistency}
\label{sec:amati}

Observationally, more energetic GRBs tend to be harder. This is 
manifested as the correlations $E_p \propto E_{\gamma,iso}^{1/2}$ 
\citep{amati02} and $E_p \propto L_{\gamma}^{1/2}$ 
\citep{weigao03,yonetoku04}
with large scatter. The relations are also found to apply for 
different emission episodes within the same burst 
\citep{liang04,ghirlanda09}\footnote{Some arguments have been
raised to show that the global Amati relation is a pure 
observational selection effect\citep{nakarpiran05,band05,butler07},
but the fact that the correlation also exists internally in 
individual GRBs suggest a physical link between $E_p$ and $L_\gamma$.
}. Inspecting Eq.(\ref{Epsyn-IS}), 
interpreting this correlation within the IS synchrotron emission
model requires that $R_{14}$ does not vary significantly among 
bursts, i.e. the internal shock 
radius is insensitive to the GRB luminosity.
Inspecting Eq.(\ref{RIS}), this suggests that GRBs all share a
similar Lorentz factor regardless of their luminosities 
\citep{zhangmeszaros02c}.
Furthermore, increasing the average $\Gamma$ (and hence 
$\Gamma_s$) tends to make a GRB softer (a larger $R_{14}$),
this is in contrast to the naive expectation that 
high-$\Gamma$ GRBs tend to be harder. Recently, \cite{liang10}
discovered a tight correlation between $\Gamma$ and $E_{\gamma,iso}$
based on the deceleration signature of a sample of GRBs with known
redshift, namely $\Gamma \propto E_{\gamma,iso}^{0.27}$.
Taking the trivial proportionality of $L_\gamma \propto
E_{\gamma,iso}$, this gives $R_{\rm IS} \propto \Gamma^2 \propto 
L_\gamma^{0.54}$ and  $E_p \propto E_{\gamma,iso}^{-0.04}$ \citep{liang10},
which is far from the observed Amati relation. This is another
difficulty of the IS model.

\subsubsection{The missing bright photosphere problem}

In order to develop strong internal shocks, the composition of the 
GRB ejecta has to be baryonic, i.e. $\sigma \ll 1$. Such a baryonic
outflow is believed to be accelerated from a fireball that initially 
carries most of its energy in thermal form and converts
this energy to kinetic form during the acceleration process
\citep{paczynski86,goodman86,shemi90}. As the fireball becomes
transparent, a bright thermal emission component is expected to
leak out from the fireball photosphere, which forms a distinct
thermal emission component in the GRB spectrum. Such a thermal
component is predicted bright enough to be usually detectable 
along with the IS non-thermal 
emission component \citep{meszarosrees00,meszaros02b,peer08}. In 
the past, since the detector bandpass (e.g. for BATSE and Swift BAT)
was not wide enough, there have been several speculations regarding 
this thermal component. The two leading possibilities are: 
(1) the observed Band-function is the non-thermal component powered by
internal shocks, and the photosphere component is either below
or above the detector energy band; (2) the observed Band function 
is the superposition of a thermal component and a non-thermal 
component \citep{ryde05,ryde09}. 
The excellent observational data of GRB 080916C with Fermi 
(thanks to the broad band coverage of Fermi GBM and LAT) 
show no evidence of deviation of the Band-function spectrum
both below and above $E_p$. This immediately rules out these 
possibilities (1) and (2). 
If one accepts that the observed non-thermal emission from
GRB 080916C is the emission from the internal shock, a profound 
question would be ``where is the thermal component?''.
In fact, using the observed minimum variability time scale revealed by 
the GRB 080916C data, the predicted photosphere thermal emission
component is more than one order of magnitude brighter than the  
observed flux. This raises a severe problem to the fireball acceleration
scenario and the straightforward IS model \citep{zhangpeer09}.
This is the ``missing bright photosphere'' problem of the IS model. 

A plausible scenario is to interpret the entire Band spectrum
as the emission from the photosphere. This requires energy
dissipation below and above the photosphere (defined by
Thomson scattering optical depth being unity). This model is
discussed in detail in Sect.\ref{sec:MHD} below.

\subsubsection{Other variants of the IS model}

There are two variants of the IS model that invoke different
radiation mechanisms other than synchrotron emission. 
These variants all suffer
from the same efficiency and fast cooling problems, but may
introduce extra ingredients to confront other problems/issues
discussed above. The first one
is to invoke synchrotron self-Compton (SSC) as the mechanism for GRBs 
\citep{panaitescu00b,kumarpanaitescu08,racusin08,kumarnarayan09}.
The allowed parameter space for SSC is larger than synchrotron
to interpret the GRB data \citep{kumarmcmahon08}. However,
the SSC model suffers from the following drawbacks. First, it
generally predicts a bright prompt optical emission component.
Although this is consistent with GRB 080319B \citep{racusin08},
prompt optical emission data of other GRBs are generally 
consistent with the extrapolation of the
gamma-ray spectrum to the optical band \citep{shen09}.
Second, a dominant SSC component inevitably predicts higher
order SSC components, which greatly increases the energy
budget of GRBs \citep{derishev01,kobayashi07,piran09}.
Third, the synchrotron/SSC model may not reproduce the much more
variable gamma-ray light curve (than the optical light curve) observed 
in GRB 080319B \citep{resmi10}.
Finally, $E_p$ in the SSC model very sensitively depends on
the Lorentz factor of the electrons ($E_p \propto \gamma_{e,p}^4$),
so that a small dispersion of the $\gamma_{e,p}$ distribution
gives a very wide range of $E_p$ distribution \citep{zhangmeszaros02c}.
This makes the parameters more contrived to account for the
observed narrow clustering of $E_p$ for bright BATSE GRBs 
\citep{preece00}.

Another IS model variant is to invoke ``jitter'' emission as
the radiation mechanism \citep{medvedev00}. 
In this scenario, the magnetic field in the
emission region has a random configuration, with a coherent length
scale $\lambda_B \ll R_B=\gamma_e m_e c^2/eB'$, the Larmor radius.
Electrons in such a field hardly make one gyration before the field
direction changes. As a result, the typical frequency of the
radiation spectrum is no longer related to the strength of the
field, but is related to the coherent length, i.e.
\begin{equation}
E_p \sim \hbar \Gamma \gamma_{e,p}^2 \frac{c}{\lambda_B} (1+z)^{-1},
\label{Epjitter}
\end{equation} 
which is by definition much larger than $E_p$ in the synchrotron
emission model (Eq.[\ref{Epsyn}]) given the same $\gamma_{e,p}$.
This eases the constraint discussed in \S\ref{sec:enumber}.
However, since there is no prediction on $\lambda_B$ from the first
principle (and it is not clear whether such a characteristic scale
exists), it is not easy to assess how $E_p$ depends on model
parameters. In particular, the correlation $E_p \propto L_\gamma^{1/2}$
established in the synchrotron model (Eq.[\ref{Epsyn-IS}]) is no 
longer straightforward to justify. Recently, \cite{sironi09b} 
synthesized the particle spectrum from PIC simulations and concluded
that the spectrum is entirely consistent with synchrotron radiation
in the magnetic fields generated by Weibel instability.
The ``jitter'' regime is recovered only when one artificially 
reduces the strength of the electromagnetic fields. 

\subsection{The electromagnetic model}\label{sec:EM}

In another extreme, i.e. $\sigma \gg 1$, a so-called
ElectroMagnetic model (EM model) \citep{lyutikov03,lyutikov06b}
has been proposed. In the high-$\sigma$ regime, the comoving
Alfv\'en speed is close to speed of light, which can be 
written as \citep{jackson75} 
\begin{equation}
V'_{\rm A} = \frac{c V'_{\rm A,NR}}{(c^2+{V'_{\rm A,NR}}^2)^{1/2}}~,
\end{equation}
where 
\begin{equation}
V'_{\rm A,NR}= \frac{B'}{\sqrt{4\pi \rho'}} = \sqrt{\sigma} c
\end{equation} 
is the comoving Alfv\'en speed in the non-relativistic regime
(noticing the definition of  $\sigma$ in Eq.[\ref{sigma}]),
one can write the Lorentz factor of the comoving Alfv\'en wave as
\begin{equation}
\gamma'_{\rm A} = (1+\sigma)^{1/2}~.
\label{gammaA}
\end{equation}
For a cold plasma (i.e. the comoving thermal energy density much
smaller than the rest mass energy density, and hence, the magnetic
energy density), gas sound speed is much less than the Alfv\'en 
speed. Therefore $\gamma'_A$ is also approximately the comoving 
Lorentz factor of other magnetoacoustic waves. 
The EM model \citep{lyutikov03} applies to the ``sub-Alfv\'enic''
regime, i.e. $\Gamma < \gamma'_A = (1+\sigma)^{1/2}$,
or roughly 
\begin{equation}
\sigma > \sigma_{c} \equiv \Gamma^2-1 \sim 10^6 \Gamma_3^2~
\label{sigmac}
\end{equation}
at the deceleration radius.
According to this model, the Poynting flux is dissipated through
electromagnetic current-driven instabilities. The model has some 
novel features that are not shared by the fireball IS model. 
For example, the emission radius where strong magnetic dissipation
occurs is $R \sim 10^{16}$ cm, much larger than that of internal
shocks defined by the minimum variability time scales
(Eq.[\ref{RIS}]). This large radius is consistent with independent
constraints on GRB emission site using different methods 
\citep{lyutikov06,kumar07,shen09,gupta08,zhangpeer09}.
The EM model also justifies a ``structured'' jet with energy
per unit solid angle dropping with angle as $E(\theta) \propto
\theta^{-2}$, as has been invoked in some GRB phenomenological
models \citep{meszaros98,rossi02,zhangmeszaros02b}.

There are however two major issues related to the EM model.
The first one is regarding its extremely high-$\sigma$ value
at the deceleration radius.
\cite{lyutikov03} argued a force-free flow by invoking 
$\sigma \sim 10^9$ beyond the photosphere. However, considering
more realistic models to launch a magnetized outflow from the
central engine would lead to a range of $\sigma$ values that
are below $\sigma_c$ \citep[e.g.][]{spruit01}. At the extreme,
even the outflow consists of cold (pressureless) matter
accelerated exclusively by the magnetic field, the achievable
$\sigma$ is at most $\sigma_c$ \citep{michel69}. Such a maximally
magnetized outflow is achievable only at a specified geometry,
i.e. a purely radially expanding outflow. For more general
geometric configurations, the achievable $\sigma$ should be
below $\sigma_c$, and could be as low as $\sigma \leq 1$
\citep{begelman94,spruit01}. If one considers the ``collapsar''
scenario of long GRBs \citep{woosley93,macfadyen99}, a PFD 
jet launched from the central engine needs to penetrate 
through the stellar envelope. The $\sigma$ value would be 
further degraded. Recent numerical simulations 
\citep{tchekhovskoy08,nagataki09} suggest that
a high-$\sigma$ flow at the base of the flow would become
a moderate-$\sigma$ flow as it escapes the star 
\citep[cf.][]{wheeler00}. 

Another issue of the EM model is related to GRB
variability time scales. According to the EM model, the 
GRB central engine is assumed to launch a magnetic bubble,
which expands and exits the star without degrading $\sigma$,
and magnetic energy is finally dissipated by the current-driven
instability at a large radius $R\sim 10^{16}$ cm
\citep{lyutikov03}. The observed rapid variability of the
GRB light curves is interpreted as due to emission from some 
``fundamental emitters'' that are moving relativistically
with respect to the dissipation region, which itself is
moving relativistically with respect to the observer.
A related model invoking relativistic, possibly magnetic turbulence
was recently proposed by \cite{narayan09} and \cite{kumarnarayan09}. 
Within such a picture, the observed variability time scale does 
not reflect the behavior of the GRB central engine. 
As discussed earlier in \S\ref{sec:IS}, 
the connection between
at least the slow variability time scales with GRB central 
engine activity is strongly supported by the data \citep{liang06}, 
especially in view of the discovery of X-ray flares that extend
central engine activity to later times 
\citep{burrows05,chincarini07,falcone07}.
Attributing the observed variability to random motion of the
relativistic ``fundamental emitters'' in the emission region
\citep{lyutikov03} would then require arguing that the
``$T_0$-reset'' effect of the last pulse and X-ray flares 
\citep{liang06} is purely due to a chance coincidence.

\subsection{MHD and dissipative photosphere models}\label{sec:MHD}

If the magnetized ejecta can be approximated as a fluid and 
described by magnetohydrodynamics (MHD) \citep{spruit01}, the GRB 
models are in the MHD regime\footnote{\cite{lyutikov06b} defines
this regime as $1< \sigma < \sigma_{c} \equiv \Gamma^2-1 \sim 10^6 
\Gamma_3^2$, opposed to the sub-Alfv\'enic regime invoked by the
EM model. He also relates the 
``super-Alfv\'enic'' and ``sub-Alfv\'enic'' regimes to whether
or not a shock can exist in the ejecta. Such a connection is 
relevant only for 1D ejecta-medium interaction. A more relevant
condition to discuss shock formation condition in the high-$\sigma$
regime is whether the magnetized ejecta encounters a stronger 
pressure than its own magnetic pressure \citep{zhangkobayashi05,mizuno09}. 
For collisions between two shells with the same $\sigma$ value,
internal shocks can develop even for a small relative Lorentz 
factor between the two shells. For an expanding shell with small 
width (thin shell) and conical geometry, the reverse shock cannot
form for $\sigma > (10-100)$ \citep{zhangkobayashi05}.}. 
Within this regime, outflows launched from the central engine
would carry a dynamically important magnetic field, 
but it can be still approximated
as a MHD fluid\footnote{Strictly speaking, the outflow may not
satisfy the MHD condition at all radii. Since the particle density
$n \propto R^{-2}$ decays with radius faster than the 
Goldreich-Julian density $n_{_{\rm GJ}} \propto R^{-1}$, the MHD condition
would break above a critical radius $R_{_{\rm MHD}}$. If this radius 
is beyond the deceleration radius ($R_{_{\rm MHD}} > R_{\rm dec}$) 
\citep{spruit01}, then the MHD condition would not break before 
afterglow sets in. However, if $R_{_{\rm MHD}} < R_{\rm dec}$, the
MHD condition breaks at $R_{_{\rm MHD}}$ and a strong magnetic 
dissipation occurs between  $R_{_{\rm MHD}}$ and $R_{\rm dec}$.
This happens if $\sigma$ is larger than a few hundred 
\citep{zhangmeszaros02c}. We do not discuss this regime in
this paper by limiting our discussion to the regime of
$1\lesssim \sigma \lesssim 100$.}. 
The $\sigma$ value may decrease with increasing radius due to
two mechanisms. First, below the photosphere, reconnection dissipation
energy is thermalized in the plasma because of the high optical depth
of the radiated photons. The thermal energy is then converted to bulk 
kinetic energy similar to the fireball acceleration process
\citep{drenkhahn02b,drenkhahn02}. 
This scenario is more relevant to the so-called ``striped wind''
magnetic field configuration with alternating magnetic polarity
(relevant to pulsar-like central engine with misaligned magnetic
and rotational axes). For helical magnetic configurations (relevant
to black hole central engine with rotational and magnetic axes 
aligned), the neighboring ordered field lines typically have
the same orientation, so that reconnection is greatly suppressed.
 Second, the Poynting flux may
be directly converted to kinetic energy without dissipation
\citep[e.g.][]{vlahakis03,komissarov09} under the pressure 
gradient of magnetic fields in an MHD flow. 
The efficiency of such a conversion is, however, subject
to uncertainties. Without an external pressure confinement, the
Poynting flux energy cannot be fully converted to kinetic
energy, and the flow can be only accelerated up to 
$\Gamma_{\rm tot}^{1/3}$, where $\Gamma_{\rm tot}$ is the Lorentz 
factor for total conversion \citep{beskin98}. This was the
origin of the well known ``$\sigma$ problem'' of the Crab nebula.
The jet acceleration can be efficient with an external confinement,
which is relevant to GRBs in the collapsar scenario 
\citep[e.g.][]{tchekhovskoy08}. In any case, the pressure gradient 
has to be small enough for efficient acceleration to happen
\citep{lyubarsky09}. Overall, if the jet has $\sigma_0 
\gg 1$ at the central engine, the $\sigma$ value at the GRB 
emission region can range from $\sigma \leq 1$ to 
$\sigma \leq \sigma_0$, depending upon how efficient
the conversion from Poynting flux to kinetic flux would be. 
\cite{lyubarsky09} showed that it is difficult
to have a completely matter dominated jet (say, $\sigma <0.1$)
at the GRB emission radius for a dissipation-less jet with
an initial high $\sigma$, even if the most efficient conversion
occurs (see also Levinson 2010). 

In the past, the GRB prompt emission models in the MHD regime
\citep{spruit01,drenkhahn02b,drenkhahn02,vlahakis03,giannios08}
have been ``semi-fireball'' like. It is generally assumed that
an initial high $\sigma$ is reduced to $\sigma <1$ before the
jet reaches the internal shock radius, 
so that the traditional IS model can still apply at $R_{\rm IS}$.
Such a model shares essentially the same advantages and drawbacks
of the IS model. The data of GRB 080916C give a strong constraint
on such a model: the lack of a photosphere component requires that
either $\sigma \gg 1$ at the photosphere, or $\sigma < 1$ already
at the photosphere. For the former, the magnetic-to-kinetic energy
transition efficiency is low. One cannot have matter-dominated
internal shocks. For the latter,
the magnetic acceleration proceeds in a ``cold'' and efficient way,
which is difficult to achieve theoretically. The external 
confinement of the high-$\sigma$ flow (which is required
for efficient magnetic acceleration) would inevitably cause heating
of the jet to result in a bright photosphere, which is constrained
by the data of GRB 080916C.
Even if the flow could become baryon-dominated through
magnetic acceleration without a hot photosphere component, the 
internal shock efficiency is so low that only a small
fraction of energy (say $\sim 10\%$) is converted to gamma-ray 
radiation. This greatly increases the total energy budget of the
outflow, and demands an even higher $\sigma$ at the central engine
to begin with (from the missing bright photosphere argument). This makes 
it even more difficult to efficiently convert Poynting flux energy 
to kinetic energy. 

A plausible scenario is to argue that the observed spectrum
is dominated by the emission from the photosphere itself,
and that the internal shock 
emission is too weak to be detected. Within such a scenario,
there is no ``missing photosphere'' problem, since the observed
spectrum is the photosphere emission itself.
One then needs to argue that
the entire non-thermal GRB spectrum is the reprocessed 
photosphere thermal emission. This requires significant energy 
dissipation below and above the photosphere 
\citep{thompson94,rees05,thompson06,thompson07,ghisellini07c,giannios08,beloborodov09,lazzati10}.
Within the MHD models, the dissipation source is continuous 
magnetic reconnection (for the striped-wind geometry). Such a
model can be also developed for a neutron-rich baryonic 
flow with $\sigma \ll 1$, in which neutron decay 
provides a source of continuous heating in the jet 
\citep{beloborodov09}. By properly considering Compton upscattering of
photosphere thermal photons, a hard power law tail can be produced
\citep{beloborodov09,lazzati10}. This model can also naturally 
interprets GRB variability in terms of the central engine energy
injection history. The radiative efficiency is also naturally
high \citep{lazzati09}. Confronting the GRB 080916C data, however,
this model faces two difficulties. 

First, the maximum photon energy
in the dissipative photosphere model is $\sim 0.1-1$ GeV 
\citep{beloborodov09,lazzati10}. The detection of 13.2 GeV photon
associated with one of the GBM pulses in GRB 080916C 
(which corresponds to 70.6 GeV
rest-frame) then disfavors the photosphere origin of gamma-ray
emission in GRB 080916C. It has been argued that the GeV emission
of LAT GRBs including GRB 080916C may be of the external shock
origin \citep{kumar09,ghisellini09}. However, the required
forward shock parameters are extreme, which are not easy to
accommodate within the known relativistic shock models
\citep{kumar10,li10b,piran10}. Furthermore, the LAT-band
emission and GBM-band emission generally trace each other
during the prompt emission phase 
\citep{zhang10}\footnote{This is the case even 
for GRB 090902B, which clearly shows two distinct spectral 
components.}. For GRB 080916C, the time-resolved spectral analysis
with the finest temporal resolution defined by statistics suggests
that the Band function fits well the data for every time bin
throughout the burst. The peak of the GeV emission in the
logarithmic light curve coincides with the second peak of the 
GBM light curve.  All these strongly suggest that the entire 
emission of GRB 080916C is from the same emission region,
and is therefore likely 
of an internal origin (see Zhang et al. 2010 for more detailed 
discussion). The long term GeV emission, on the other hand, 
decays slower than the MeV emission. It may be dominated by
a different emission component (e.g. external shock). 

Second, although the high energy photon spectral index above 
$E_p$ can be well reproduced in the dissipative photosphere model
by Compton upscattering \citep{beloborodov09,lazzati10},
the low energy photon spectral index below $E_p$ is typically
much harder than what is observed. For up-Comptonization of a
blackbody spectrum (e.g. within the context of soft gamma-ray 
repeaters and anomalous X-ray pulsars), the Rayleigh-Jeans low 
energy spectral index (Band-function index $\alpha=+1$) is hardly
affected \citep{nobili08}). By invoking dissipation below the
photosphere (e.g. neutron heating), this index can be modified
to $\alpha=+0.4$ \citep{beloborodov09}, still much harder than
the observed typical value $\alpha \sim -1$ 
\citep{abdo09,zhang10}. One possibility would be to assume
that the observed low energy spectrum is the superposition of
the photosphere emission of many shells 
\citep[e.g.][]{toma09,toma10}. Although this
is not impossible, it requires properly arranging the luminosity
and Lorentz factor of many mini-shells to mimic a typical
$\alpha=-1$ Band spectrum. A detailed time-resolved spectral
analysis of GRB 080916C and other ``Band-only'' GRBs suggests
that the low energy photon index $\alpha$ remains essentially
unchanged as the time bin becomes progressively small
\citep{zhang10}.
This at least disfavors the possibility that the observed
time-integrated spectrum is the temporal superposition of
the photosphere emission of many shells. 
One plausible solution would be to introduce synchrotron
and synchrotron self-Compton (SSC) of the electrons. In the 
MHD model, if continuous magnetic heating is operating,
the SSC component would produce a typical Band spectrum in the
MeV range, if most energy dissipation happens in the Thomson-thin
region above the photosphere \citep[e.g. Fig. 2 of][]{giannios08}.
Such a model predicts a bright optical emission component
(synchrotron). Although this is consistent with the case of 
GRB 080319B \citep{racusin08}, it is inconsistent with most
of the prompt optical emission data or upper limits, which are 
consistent with or below the extrapolation of gamma-ray spectrum 
to the optical band \citep{shen09}.

\subsection{Summary}\label{sec:summary}

In summary, the new Fermi LAT data of GRB 080916C raises
a challenge to the traditional fireball IS model, which
has some advantages and drawbacks. The EM model invokes too
high a $\sigma$ at the deceleration radius, which may not be
achieved in nature. We consider the moderate-$\sigma$ MHD regime 
a relevant regime for GRB outflows. However, the current 
semi-fireball MHD models or the dissipative photosphere models are 
not ideal to adequately interpret GRB 080916C. The ICMART model
proposed in this paper is meant to inherit the merits of the
existing models and to overcome their drawbacks. 

Since most LAT GRBs are found similar to GRB 080916C to show 
``Band-only'' time-resolved spectra \citep{zhang10}, 
we argue that the proposed model would be relevant to most GRBs.

\section{The ICMART model}\label{sec:ICMART}

\subsection{Basic assumptions}

The Internal-Collision-induced MAgnetic Reconnection and 
Turbulence (ICMART) model is based on the following two 
assumptions:
\begin{enumerate}
\item The GRB central engine is intermittent in nature,
which ejects an unsteady outflow with variable Lorentz factors 
and luminosities but a nearly constant degree of magnetization. 
Approximating the unsteady outflow
as some discrete shells with variable Lorentz factors, these 
magnetized shells collide
with each other at the conventional internal shock radius
$R_{\rm IS}$ (Eq.[\ref{RIS}]).
\item The GRB central engine is highly magnetized, which ejects
a high-$\sigma$ flow from the base of the central engine. 
Although various mechanisms (e.g. magnetic acceleration and 
reconnection) may reduce $\sigma$ as the outflow streams
outwards, the ejecta is still moderately magnetized in the
GRB emission region, with $\sigma$ ranging from $1 < \sigma
\lesssim 100$.
\end{enumerate}

In an ideal magnetically driven MHD jet, magnetic energy is converted
into kinetic energy during the expansion.  Usually it is assumed 
that $\sigma_0 \gg 1$ at the central engine with $\Gamma_0 \sim 1$.
The magnetization parameter $\sigma$ then decreases with radius as 
$\Gamma$ increases with radius 
\citep{beskin98,vlahakis03,komissarov09,lyubarsky09,granot10}.
Without external pressure confinement, the early acceleration phase
reaches a Lorentz factor $\Gamma \sim \sigma_0^{1/3}$, with remaining
magnetization $\sigma \sim \sigma_0^{2/3}$. Beyond this radius 
acceleration is slow and weakly dependent on $R$, e.g. $\Gamma
\propto R^{1/3}$, and $\sigma \propto R^{-1/3}$ \citep{granot10}.
If $\sigma_0$ remains roughly constant, one expects a nearly
constant $\sigma$ at large radii, but in the meantime does not 
expect a large variation in $\Gamma$. We note, however, that near
the rapidly-rotating magnetized central engine, efficient acceleration
of ejecta may occur below the light cylinder thanks to direct 
electric field accelerate mechanisms (e.g. frame-dragging-induced
deviation from the force-free condition, Muslimov \& Tsygan 1992
within the context of pulsars). As a result, at the light cylinder 
(effectively the base of central engine), the ejecta may have 
$\Gamma_0 \gg 1$. The rapid variation of the jet power (e.g. due to 
variable accretion rate or variable spindown rate of the central 
object) may lead to a variable $\Gamma_0$ and somewhat variable 
$\sigma_0$ at the central engine. Considering later magnetic 
acceleration, one may get a large $\Gamma$ variation within the 
outflow, with $\sigma$ not significantly varying at a same
radius. Our above two assumptions may be then fulfilled.

\subsection{Basic parameters of the GRB ejecta plasma}

In preparation of proposing the model, it is informative
to summarize the basic parameters of  GRB ejecta.
The gamma-ray emission radius $R$ can be in principle in the 
range of $10^{11}-10^{17}$ cm (bracketed by the photosphere radius
and the external shock deceleration radius). 
As will become evident later, several collisions are needed to
trigger an ICMART event, so that one would have
$R_{\rm ICMART} \geq R_{\rm IS}$. In the following, in
order to make differentiation with the IS model, we will
normalize all radii relevant for the ICMART model to
$R=10^{15}~{\rm cm}~ R_{15}$, and denote $R$ as 
$R_{_{\rm ICMART}}$ whenever relevant.

\begin{enumerate}
\item {\em Length scales}:
At $R_{_{\rm ICMART}}$, the ``thickness'' of the ejecta 
is related to the radius $R$\footnote{This approximation is based
on the assumption of shell-spreading. For high-$\sigma$ shells, the
spreading condition may be more stringent \citep{granot10}. The 
shell width is then related to the duration of the central engine, not a 
function of $R$. In any case, the following estimates are valid
to order of magnitude.}, i.e.
\begin{equation}
\Delta'=\frac{R}{\Gamma} \simeq 3.2\times 10^{12}~{\rm cm} R_{15}
\Gamma_{2.5}^{-1}
\label{Del}
\end{equation}
in the comoving frame, and
\begin{equation}
\Delta=\frac{R}{\Gamma^2} \simeq 10^{10}~{\rm cm} R_{15}
\Gamma_{2.5}^{-2}
\end{equation}
in the rest frame of the central engine (lab frame).
Assuming a conical jet with the opening angle $\theta_j$, the cross
section radius of the emission region is
\begin{equation}
R_\theta = R \theta_j = 8.7\times 10^{13} \left(\frac{\theta_j}
{5^{\rm o}}\right) R_{15}~{\rm cm}
\label{Rtheta}
\end{equation}
in both the lab frame and the comoving frame. Typically one has
$R_\theta \gg \Delta' \gg \Delta$. GRB ejecta are therefore 
also considered as ``flying pancakes''.
\item {\em Plasma number density}:
In a conical jet, density drops with radius as $ R^{-2}$.
For a hydrogen ejecta with a total ``wind'' luminosity $L_w$, Lorentz 
factor $\Gamma$, and magnetization parameter $\sigma$, the comoving 
ejecta proton number density is
\begin{eqnarray}
n'_p & = & \frac{L_w}{4\pi (1+\sigma) R^2 \Gamma^2 (m_p+Y m_e) c^3}
\nonumber \\
& \simeq & 1.8\times 10^7 ~{\rm cm^{-3}} L_{w,52} \Gamma_{2.5}^{-2}
R_{15}^{-2} m^{-1}(1+\sigma)_1^{-1}~,
\label{np'}
\end{eqnarray}
where $m=1+Y m_e/m_p \sim 1$ if the lepton (pair) multiplicity parameter
$Y \ll m_p / m_e$. In the rest frame of the central engine,
the ejecta proton number density is 
\begin{equation}
n_p = \Gamma n'_p \simeq 5.6 \times 10^9 ~{\rm cm^{-3}} 
L_{w,52} \Gamma_{2.5}^{-1}
R_{15}^{-2} m^{-1}(1+\sigma)_1^{-1}~.
\end{equation}
The electron number (or more generally the lepton number) densities are
\begin{equation}
n'_e = Y n'_p, ~~~~~ n_e = Y n_p~.
\label{ne'}
\end{equation}
\item {\em Magnetic field strength}:
The comoving magnetic field strength is
\begin{eqnarray}
B' & = & \left( \frac{L_w}{\Gamma^2R^2 c} \frac{\sigma}{1+\sigma}
\right)^{1/2}
\nonumber \\
& \simeq & 1.8 \times 10^3 ~{\rm G} \left(\frac{\sigma}{1+\sigma}
\right)^{1/2} L_{w,52}^{1/2} \Gamma_{2.5}^{-1} R_{15}^{-1} ~.
\label{B'}
\end{eqnarray}
In the lab frame, the magnetic field strength is
\begin{equation}
B = \Gamma B' \simeq  5.8 \times 10^5 ~{\rm G} \left(\frac{\sigma}{1+\sigma}
\right)^{1/2} L_{w,52}^{1/2} R_{15}^{-1} ~.
\end{equation}
This ${\bf B}$ field is accompanied by an induced ${\bf E}=-{\bf V
\times B}$ field for an ideal MHD fluid.
\item {\em Collisional mean free path and time scale}:
For Coulomb collisions, the strong collision radius may be defined
by $e^2/r_{col} \sim k T$ so that $r_{col} \sim e^2/k T
\sim (1.7\times 10^{-3}~{\rm cm})/T$. Here $kT$ denotes more generally
the typical energy of the particles (which are not necessarily in thermal
equilibrium). The comoving collision mean 
free path of electrons can then be estimated as
\begin{eqnarray}
l'_{e,col} &  = & (n'_e \pi r_{col}^2)^{-1}  \simeq 
6.5\times 10^{17}~{\rm cm} \nonumber \\
& \times & 
L_{w,52}^{-1} \Gamma_{2.5}^2 R_{15}^2 m Y^{-1}(1+\sigma)_1 
T_{e,10}^2~.
\end{eqnarray}
In order to have the plasma in the ``collisional'' regime, one 
needs to require $l'_{e,col} < \Delta'$. This is translated into
$T_e < T_{e,c} \equiv 2.2 \times 10^7 ~{\rm K}~ L_{w,52}^{1/2} 
\Gamma_{2.5}^{-3/2} R_{15}^{-1/2} m^{-1/2} Y^{1/2} 
(1+\sigma)_1^{-1/2} $. Downstream of the GRB internal shocks, 
the effective proton temperature is of the order of 
$T_{p} \sim (\Gamma_{ud}-1) m_p c^2/k \sim 1.1\times 10^{13} ~{\rm K}~
(\Gamma_{ud}-1)$ (where $\Gamma_{ud}$ is the relative Lorentz factor
between upstream and downstream), 
and the electron temperature is even higher
by a factor of $(\epsilon_e/\epsilon_p)(m_p / m_e)$. As a result,
GRB shocks must be ``collisionless''. In the upstream, 
the temperature can be lower, but since the flow is relativistic,
any inhomogeneity in the velocity field would result in ``heating''
in the flow. Other processes (magnetic reconnection, neutron decay,
etc) would also enhance heating, so that the electron temperature 
may be maintained to be significantly 
above $T_{e,c}$. For a reasonable estimate, the electrons
in a relativistic flow would have at least a relativistic temperature
$T_e \sim m_e c^2/ k
= 5.9\times 10^9$ K. The collision time scale can be estimated as 
$\tau'_{col}=l'_{e,col} /v'_e$. For a non-relativistic temperature, 
one has $v'_e=(2kT_e/m_e)^{1/2}$, so that the comoving collision time
$\tau'_{col,NR} = {m_e^{1/2} (kT_e)^{3/2}}/{(\sqrt{2}\pi e^4 n'_e)}$.
In the GRB ejecta (even without shock heating), it is very possible 
that the electrons have a comoving relativistic temperature, so that 
$v'_e \sim c$. The comoving collisional time is therefore
\begin{eqnarray}
\tau'_{col,R} & = & \frac{l'_{e,col}}{c}\simeq 2.2\times 10^7~{\rm s}
\nonumber \\
& \times & L_{w,52}^{-1} \Gamma_{2.5}^2 R_{15}^2 m 
Y^{-1}(1+\sigma)_1 T_{e,10}^2~,
\end{eqnarray}
which is $\gg t'_{dyn}= R/\Gamma c \sim 110~{\rm s} R_{15} \Gamma_{2.5}^{-1}$,
the comoving dynamical time scale. This again suggests that ejecta is 
collisionless even without strong shock heating.
\item {\em Gyroradii and gyrofrequencies}:
Another relevant length scale is the particle gyro-radius in magnetic 
fields. Without direct collisions, the GRB ejecta can be still approximately 
described as a ``fluid'' macroscopically. This is because particles are 
interacting with each other through magnetic fields microscopically.
Fundamentally the smallest length scale is defined by particle gyration.
The comoving frame gyro(cyclotron)-radii are
\begin{equation}
r'_{B,e} = \frac{\gamma_e m_e c^2}{e B'} \simeq 0.93~{\rm cm}~\gamma_e
L_{w,52}^{-1/2} \Gamma_{2.5} R_{15} \left(\frac{1+\sigma}{\sigma}
\right)^{1/2}
\label{rce}
\end{equation}
for electrons (where $\gamma_e$ is the electron Lorentz factor), and
\begin{equation}
r'_{B,p} = \frac{\gamma_p m_p c^2}{e B'} \simeq 1.7\times 10^3~{\rm cm}~
\gamma_p L_{w,52}^{-1/2} \Gamma_{2.5} R_{15} \left(\frac{1+\sigma}{\sigma}
\right)^{1/2}
\label{rcp}
\end{equation}
for protons (where $\gamma_p$ is the proton Lorentz factor).
Given the typical $\gamma_e$ values to interpret the GRB emission
(Eq.[\ref{gammaep}]) and the related $\gamma_p$ (which is typically
smaller by a factor of $\epsilon_p m_e/\epsilon_e m_p$), 
both radii are $\ll \Delta'$. If the typical viscous length scale is
not much larger than the gyration radius, the fluid description 
of the GRB plasma is justified. The comoving gyrofrequencies are
\begin{equation}
\omega'_{B,e} = \frac{eB'}{m_e c} \simeq 3.2\times 10^{10}~{\rm s^{-1}}
L_{w,52}^{1/2} \Gamma_{2.5}^{-1} R_{15}^{-1} \left(\frac{\sigma}{1+\sigma}
\right)^{1/2}
\label{omegace}
\end{equation}
for electrons, and 
\begin{eqnarray}
\omega'_{B,p} & = & \frac{eB'}{m_p c}=\omega'_{B,e} \frac{m_e}{m_p}
\nonumber \\
& \simeq & 1.7 \times 10^7~{\rm s^{-1}}
L_{w,52}^{1/2} \Gamma_{2.5}^{-1} R_{15}^{-1} \left(\frac{\sigma}{1+\sigma}
\right)^{1/2}
\end{eqnarray}
for protons. Both are much larger than the inverse of the comoving 
dynamical time, i.e. $(R/\Gamma c)^{-1} \sim 9.5\times 10^{-3}
~{\rm s^{-1}}~\Gamma_{2.5} R_{15}^{-1}$, again justifying the
fluid description.
 \item {\em Plasma frequencies and plasma skin depths}: 
The comoving relativistic plasma frequencies are
\begin{eqnarray}
\omega'_{p,e} & = & \left(\frac{4\pi n'_e e^2}{\bar\gamma_e m_e} 
\right)^{1/2} \simeq 2.4\times 10^{8}~{\rm s^{-1}} \nonumber \\
& \times & \bar\gamma_e^{-1/2}
 Y^{1/2} L_{w,52}^{1/2} \Gamma_{2.5}^{-1} R_{15}^{-1}
 m^{-1/2}(1+\sigma)_1^{-1/2}
\label{omegape}
\end{eqnarray}
for electrons, and
\begin{eqnarray}
\omega'_{p,p} & = & \left(\frac{4\pi n'_p e^2}{\bar\gamma_p m_p} \right)^{1/2}
\simeq 5.5\times 10^{6}~{\rm s^{-1}} \nonumber \\
& \times & \bar\gamma_p^{-1/2}
L_{w,52}^{1/2} \Gamma_{2.5}^{-1} R_{15}^{-1}m^{-1/2} (1+\sigma)_1^{-1/2}
\end{eqnarray}
for protons. Here $\bar\gamma_e$ and $\bar\gamma_p$ denote the mean
Lorentz factor of the relativistic electron and proton gas, respectively.
The corresponding plasma skin depths are
\begin{equation}
\delta'_e=\frac{c}{\omega'_{p,e}} \simeq 130~{\rm cm}~ \bar\gamma_e^{1/2}
Y^{-1/2} L_{w,52}^{-1/2} \Gamma_{2.5} R_{15} m^{1/2} (1+\sigma)_1^{1/2}
\end{equation}
for electrons, and
\begin{equation}
\delta'_p=\frac{c}{\omega'_{p,p}} \simeq 5.4\times 10^3~{\rm cm}~ 
\bar\gamma_p^{1/2} L_{w,52}^{-1/2} \Gamma_{2.5} R_{15} m^{1/2} (1+\sigma)_1^{1/2}
\end{equation}
for protons. Physically these quantities are relevant to a
weakly magnetized ejecta with random $B$ fields ($\sigma \ll 1$). For a 
$\sigma>1$ flow with ordered magnetic fields,
the plasma oscillation frequencies and skin depths are relevant only in 
the direction parallel to the magnetic field lines. In the perpendicular
direction, the gyrofrequencies and gyroradii are more relevant.
\end{enumerate}

\subsection{Turbulent nature of the GRB ejecta}

Turbulence is believed to be ubiquitous in astrophysical systems. This is 
because the Reynold's number, the ratio between the viscous diffusion time
$\tau_\nu = L^2/\nu$ and the relative flow time scale $\tau_f=L/\delta V$, 
\begin{equation}
R_e \equiv \frac{L \delta V}{\nu}~,
\end{equation}
is $\gg 1$ (mostly because of the large $L$ involved in astrophysical
systems), where $L$ and $\delta V$ are the characteristic length and
relative velocity of the flow, and
\begin{equation}
\nu \sim c_s l
\end{equation}
is the kinematic viscosity as defined in the MHD equation
of motion
\begin{equation}
\rho \left( \frac{\partial {\bf V}}{\partial t}+{\bf V}\cdot
\nabla{\bf V}\right)=-\nabla P+{\bf J}\times {\bf B}
+\rho\nu \nabla^2 {\bf V}~,
\end{equation}
$c_s$ is sound speed, and $l$ is the mean free path of microscopic 
interactions that define the viscosity.

Since shear motion tends to distort the fluid while the
viscous term tends to smear the distortion of the fluid,
the flow would become highly distorted and turbulent 
when the Reynold's number is $\gg 1$.

Being magnetically dominated,  most kinetic motions are concentrated 
in the direction perpendicular to the magnetic field. The corresponding
perpendicular viscosity is (Spitzer 1962)
\begin{equation}
\nu_\bot = 1.7\times 10^{-2} {\rm cm^2 s^{-1}} n \ln\Lambda/(\sqrt{T} B^2),
\end{equation} 
 where 
$\ln \Lambda $ is the Coulomb logarithm. $\Lambda = 3/2e^3 
\sqrt{k^3T^3/\pi n}{\rm min}(1, \sqrt{4.2\times 10^5/T})$.
For typical parameters adopted in this paper, one has
$\nu_\perp \sim 3.7\times 10^{-4} {\rm cm^2~s^{-1}}\ll L\delta V
\sim 10^{24} {\rm cm^2~s^{-1}}$, so that $R_e \sim 10^{28} \gg 1$.
Therefore the GRB ejecta is turbulent in nature.
In the low-$\sigma$ case,  \cite{zhangw09} has shown
numerically that the GRB ejecta with a mildly relativistic
relative motion (relevant to internal shocks and late external
shock in the trans-relativistic regime) quickly turns turbulent
if a Kelvin-Helmholtz instability is triggered. In the 
high-$\sigma$ regime, due to the strong magnetic pressure
in the ordered magnetic fields, the condition to trigger
turbulence would be more demanding (see \S\ref{sec:avalanche}). 
In any case,
strong anisotropic turbulence can develop once the triggering
condition is satisfied.

For a resistive magnetized flow, another relevant dimensionless 
parameter is the
magnetic Reynold's number, the ratio between the magnetic resistive
diffusion time $\tau_{dif} =L^2/\eta$ and the flow time 
$\tau_f=L/\delta V$,
\begin{equation}
R_m \equiv \frac{L \delta V}{\eta},
\end{equation}
where $\eta$ 
is the magnetic diffusion coefficient, which is defined in
the diffusive MHD induction equation
\begin{equation}
\frac{\partial {\bf B}}{\partial t} = \nabla \times
({\bf V}\times{\bf B}) + \eta \nabla^2 {\bf B}~.
\end{equation}
For the reason described above, we adopt the perpendicular resistivity 
in a strong magnetic field (Spitzer 1962),
\begin{equation}
\eta_\bot = 1.3\times 10^{13} {\rm cm^2 s^{-1}}\frac{Z \ln \Lambda}{T^{3/2}}~.
\end{equation}
For the typical parameters adopted in this paper, one has
$\eta_\bot \sim 1~{\rm cm^2~s^{-1}}$.
This gives a huge $R_m$ number, $R_m \sim 10^{24}$.
The maximum resistivity is the ``Bohm'' diffusion, i.e.
\begin{equation}
\eta_{\rm B} \lesssim r_B V \sim r'_{B,e} c.
\label{etaB}
\end{equation}
This gives 
\begin{equation}
R_{\rm m,B} \simeq \Delta'/r'_{B,e}\simeq 3.4\times 10^{12}
\gamma_e^{-1} L_{w,52}^{1/2} \Gamma_{2.5}^{-2}
\left(\frac{\sigma}{1+\sigma}\right)^{1/2}~.
\end{equation}
This is still a large value, suggesting that
magnetic fields can be also
highly distorted and turbulent if the turbulence triggering
condition is satisfied.

Unlike the hydrodynamical turbulence that displays a Komolgorov 
solution, i.e. $dE(k) \propto k^{-5/3}$
(where $k$ is the wave number and $E(k)$ is the energy per unit
wave number), MHD turbulence is anisotropic and has different
scaling in the directions perpendicular to and along the field
line, namely $E(k_\perp) \propto k_\perp^{-5/3}$ (Komolgorov-type) 
and $E(k_\parallel) \propto k_\|^{-2}$ \citep{goldreich95,cho02}.
While the kinetic power in
the turbulence drops with $k$ quickly, the power of magnetic
fields does not drop significantly with $k$. As a result, the
``eddies'' in smaller scales are even more stretched and appear
elongated along the {\it local} magnetic field.
The MHD turbulence in the $\sigma > 1$
regime is not well studied, but qualitatively it should be 
the natural extension of the physics in the low $\beta=P_{gas}
/P_{mag}$ regime \citep{cho02}\footnote{Notice that $\beta$ is
not simply $\sigma^{-1}$, since $P_{gas}$ is the thermal pressure
of the gas, while in the definition of $\sigma$ (Eq.[\ref{sigma}]),
the rest mass energy density is invoked. Even in the very
low-$\beta$ regime ($P_{gas} \sim 0$), the $\sigma$ value can be
still below unity.}. The main extension is that the 
relative motion must be relativistic in order to distort the
magnetic field lines. This is not very demanding in a GRB,
since the outflow itself is highly relativistic. The MHD 
turbulence in a GRB ejecta is relativistic in nature.

\subsection{Magnetic reconnection in the GRB ejecta}
\label{sec:reconnection}

Reconnection plays a fundamental 
role in both laboratory and astrophysical settings, although
the full details of the process was not well understood until 
recently. The main difficulty was that in a steady state, 
the reconnection process proceeds very slowly 
\citep{sweet58,parker57}, and is not adequate 
to account for the abrupt reconnection events observed in
the lab and various astrophysical phenomena, such as solar
flares. This was merely a theoretical problem as in reality 
reconnection is indeed fast since astrophysical magnetic fields 
would be completely entangled due to turbulent motions.
In the standard non-relativistic Sweet-Parker scenario, 
two sets of field lines with opposite orientations approach 
each other and reconnect within a layer of thickness $\delta$
and length ${\cal L}$, which satisfy
\begin{equation}
\frac{\delta}{{\cal L}}=\frac{V_{in}}{V_A}= S^{-1/2}~,
\end{equation}
where the Lundquist number
\begin{equation}
S \equiv \frac{{\cal L} V_A}{\eta}
\end{equation}
is essentially the magnetic Reynold's number in the Bohm
diffusion approximation, with $\delta V$ replaced by the 
Alfv\'en speed $V_A$, and $L$ replaced by ${\cal L}$. 
Notice that for individual
reconnection events, it can well be that ${\cal L} \ll L \sim
R_\theta$ (Eq.[\ref{Rtheta}]).  In any case, $S$ is usually a 
large number, so that the Sweet-Parker process 
is an extremely slow process, i.e. 
$V_{in}=V_A S^{-1/2} \ll V_A$. A fast steady-state reconnection 
scenario was proposed by \cite{petschek64}, which invokes a 
shorter width ${\cal L}$ of the resistive layer. 
However, such a scenario is 
unstable unless the magnetic diffusion $\eta$ increases near
the X-point. Simulations with a constant $\eta$ suggests that
an initial Petschek configuration would quickly revert to the 
Sweet-Parker configuration \citep{uzdensky00}.
\cite{lazarian99} proposed that reconnections in 
magnetic fields with stochasticity can proceed
rapidly, thanks to the turbulent nature of the magnetized fluid
that both broadens the reconnection zone and allows many independent 
flux reconnection events to occur simultaneously. This turbulence 
model of 3D reconnection is confirmed by numerical simulations 
(Kowal et al. 2009), and is the basis of our discussion
of the ICMART model\footnote{We note that the study of
reconnection is an rapidly evolving field 
\citep{drake06,loureiro07,loureiro09,zenitani09,zenitani10}. 
Many questions still remain open. }.

In the presence of turbulence, a magnetic field reconnects over a local 
scale $\lambda_\|$ rather than the global scale ${\cal L}$. This is 
the scale over which a magnetic field wanders away from its original 
direction by the thickness of the Ohmic diffusion layer. Accordingly, 
it is the parameter
\begin{equation}
s \equiv \frac{\lambda_\| V_A}{\eta}
\label{tau}
\end{equation} 
rather than the Lundquist number $S$ that matters. The 
comoving local reconnection rate is given by $V'_{\rm rec,loc}
\approx V_A s^{-1/2}\ll V_A \sim c$ \citep{lazarian99}.

One may estimate the required global reconnection speed to power
a GRB. For the high-$\sigma$ flow invoked in this paper, a 
significant fraction of the magnetic field energy is converted into
radiation. This requires that the dissipated magnetic energy in a shell
with comoving width $\Delta'$ within the time scale 
$\Delta t'$ can power the observed gamma-ray luminosity,
i.e. $\Gamma^2 (B'^2/8\pi) 4\pi R^2 \Delta'
/\Delta t' \sim 10^{52} L_{\gamma,52}$. To compare against the
definition of $B'$ (Eq.[\ref{B'}]), this demands that the global
reconnection speed 
\begin{equation}
V'_{\rm rec,global}=\frac{\Delta'} {\Delta t'}
\sim \frac{L_{\gamma,52}}{L_{w,52}} \frac{1+\sigma}{\sigma}
c \sim c~.
\end{equation}
In turbulent reconnection, the global reconnection rate is 
boosted from $V'_{\rm rec,loc}$ by a factor of 
$\sim  L/\lambda_\|$,
since $\sim L/\lambda_\|$ field lines reconnect
simultaneously \citep{lazarian99}. As a result, the small
local reconnection speed $V'_{\rm rec,loc}\approx V_A s^{-1/2}$
is adequate to power a GRB if $(L/\lambda_\|) s^{-1/2}
\geq 1$, or
\begin{equation}
\lambda_\| \leq L^{2/3} \left(\frac{\eta}{c} \right)^{1/3}~.
\label{condition}
\end{equation}
For the Bohm diffusion limit (Eq.[\ref{etaB}]), and taking
$L \sim R_\theta$ (Eq.[\ref{Rtheta}]), the condition for
triggering a GRB by global turbulent reconnection can be
written as 
\begin{eqnarray}
\lambda_\| & \leq & R_{\theta}^{2/3} {r'}_{B,e}^{1/3} 
\simeq  1.9\times 10^9 ~{\rm cm} \nonumber \\
&\times & \gamma_e^{1/3} 
\left(\frac{\theta_j}{5^{\rm o}}\right)^{2/3} L_{w,52}^{-1/6}
\Gamma_{2.5}^{1/3} R_{15} \left(\frac{1+\sigma}{\sigma}\right)^{1/6}~.
\label{condition1}
\end{eqnarray}
This requires that the individual reconnection region length
scale ${\lambda_\|}$ is sufficiently small (e.g. much smaller than 
$R /\Gamma = 3.3\times 10^{12}~{\rm cm} R_{15}\Gamma_{2.5}^{-1}$, 
the typical ``observable'' scale due to relativistic beaming).
This calls for multiple collision-induced perturbations to 
trigger turbulence as we discuss below.

\subsection{Trigger of a reconnection-turbulence ``avalanche'' and 
the role of collision}\label{sec:avalanche}

We envisage internal interactions within the wind (i.e. collisions) as 
the main agent to induce turbulence. Similar to the IS picture,
numerous collisions can occur within the wind. However, most
collisions do not lead to a strong radiation signature.
In the high-$\sigma$ regime, strong shocks can still exist
as long as the ram pressure received by a magnetized shell 
is higher than the rest-frame magnetic pressure 
\citep{zhangkobayashi05}. However, if no significant 
magnetic energy dissipation occurs, the total released 
energy is at most $(1+\sigma)^{-1}$ of the total ejecta energy,
even if all the baryonic energy is converted to heat and
gets radiated away completely. This is the main reason that
radiation efficiency is low without magnetic 
dissipation \citep{zhangkobayashi05}.

\begin{figure}
\plotone{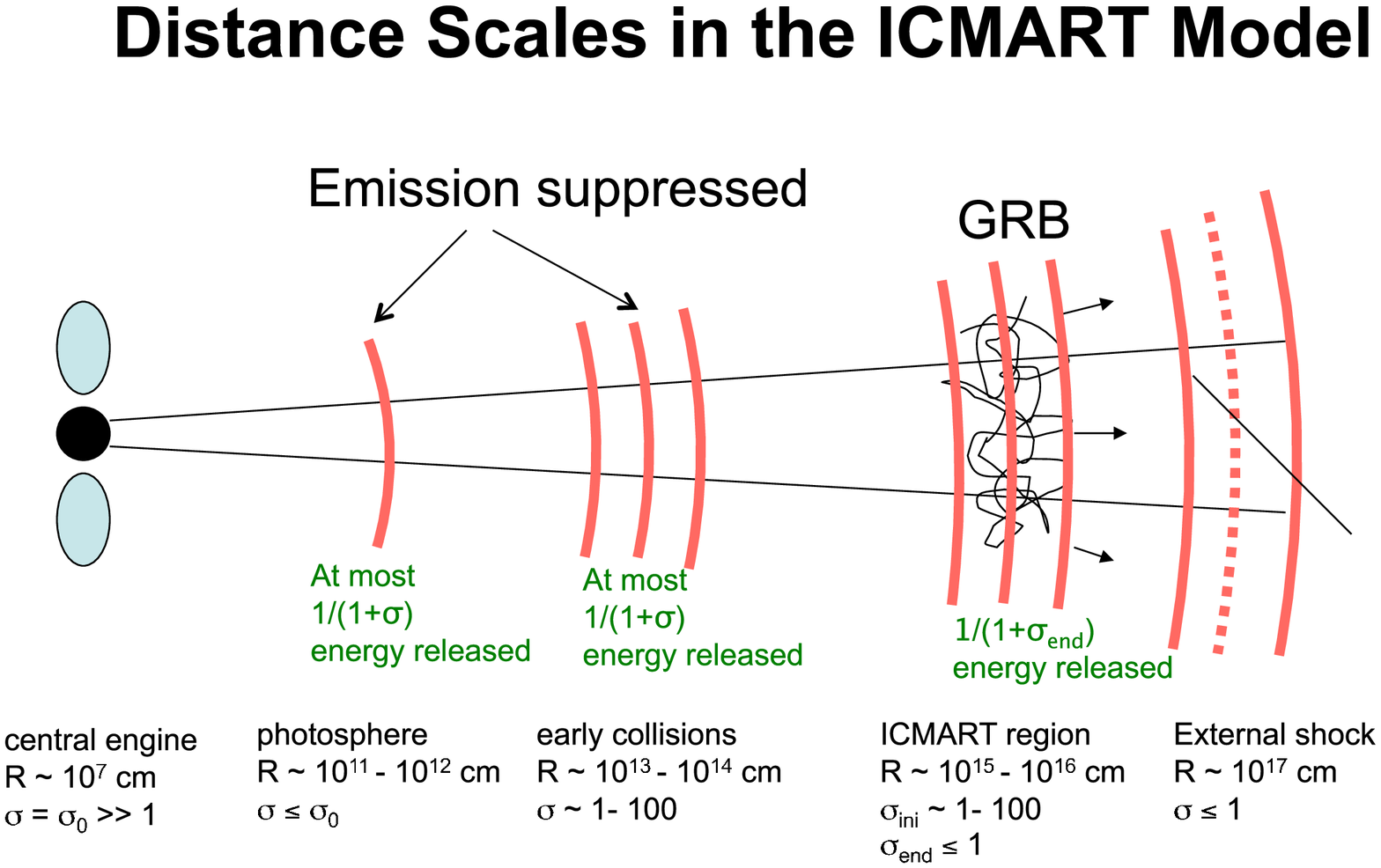}
\caption{A cartoon picture of the ICMART model. The typical distances and $\sigma$ values of various events 
are marked. }
\label{fig:cartoon}
\end{figure}

\begin{figure}
\plotone{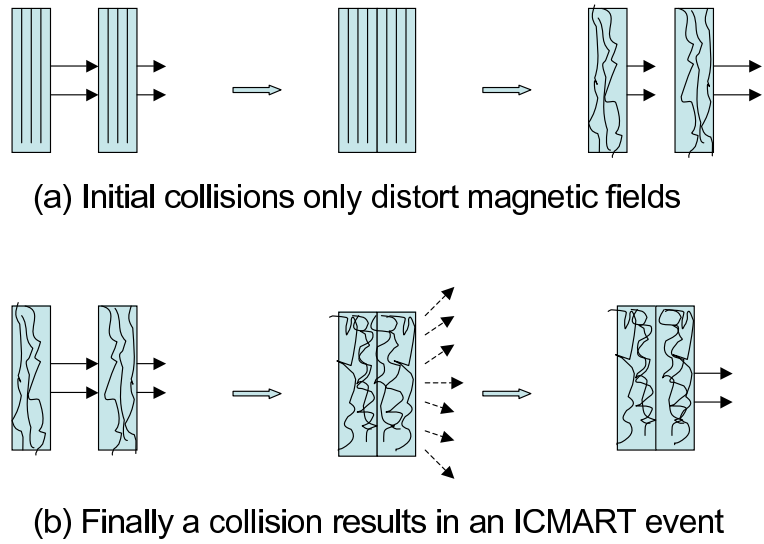}
\caption{Examples of internal collisions that mainly distort magnetic fields and result in catastrophic 
discharge of magnetic energy in an ICMART event. }
\label{fig:collisions}
\end{figure}

\begin{figure}
\plotone{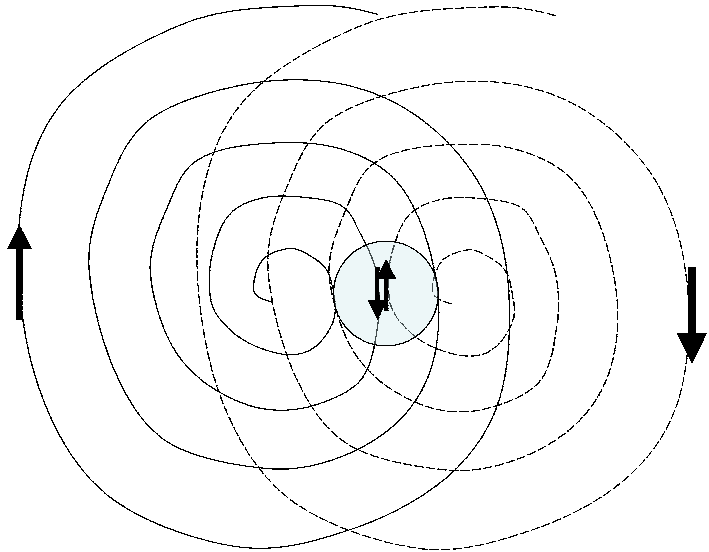}
\caption{An example of triggering an ICMART event. Magnetic field lines with opposite orientations can
approach each other and may result in fast reconnection to trigger ICMART if the two shells with
mis-aligned helical magnetic field configuration collide.}
\label{fig:trigger}
\end{figure}

In order to give rise to an efficient radiation episode, the
collision must lead to an abrupt discharge of the magnetic
field energy, so that the ending $\sigma$ value is significantly
smaller than the initial one, i.e. $\sigma_{end} \ll \sigma_{ini}$. 
This must be accompanied by an ``avalanche'' of magnetic
reconnection/turbulence events. 
A critical condition to trigger such a
run-away reconnection/turbulence avalanche has to be satisfied. 
A quantitative
description of such a critical condition is difficult, given
the complicated physics involved in magnetic reconnection and
turbulence. Nonetheless, one may envisage the following scenario
(Fig.\ref{fig:cartoon}): \\
\begin{itemize}
\item Initially, the field lines entrained in the ejecta are 
globally ordered. Since the poloidal field drops with radius more
rapidly ($\propto R^{-2}$) than the toroidal field ($\propto R^{-1}$),
the ordered field lines essentially lie in the plane of the 
ejecta front (i.e. perpendicular to the direction of motion of
the ejecta). In the literature, two types of field 
configurations have been discussed for GRBs 
\citep[e.g.][]{spruit01}: the ``striped
wind'' geometry with alternating field configuration and the
helical ``jet-like'' structure. The former is 
relevant for pulsar systems in which the magnetic axis is misaligned
with the rotational axis. For GRB systems, unless the central
object is a rapidly rotating millisecond magnetar, it is more
likely that the magnetic axis is aligned with the rotational axis
(e.g. in the black hole - torus systems). This leads to the
helical magnetic field configuration. In the following, we will
mostly focus on the helical structure, but will mention the 
striped wind geometry when relevant.

\item Internal collisions in an unsteady outflow would alter the
magnetic field configuration. Initially the field lines have a
large coherent length ($\lambda_\| \sim R_\theta$).
Since the most field lines have the same orientation, reconnection
is very difficult to occur. Even if it occurs, 
since the condition to power a GRB (Eqs.[\ref{condition}] and
[\ref{condition1}]) is not satisfied, the
reconnection process proceeds very slowly. 
Without significant magnetic dissipation, 
the collisions are essentially elastic. In any
case, the ram pressure received by each magnetized shell 
during the collisions serves as the agent to distort 
the field lines. This is because in reality, all the fields
(density, velocity and magnetic) are not uniform, and 
counter-streaming of the two fluids involved in a collision
would amplify the perturbations to make the field lines
more distorted. 
The scale of coherent length $\lambda_\|$ would 
progressively reduce as collisions continue to occur. 
As a result, the parameter $s$ would 
gradually decrease with radius (Fig.\ref{fig:collisions}).

\item The first reconnection seed requires bringing together two 
sets of field lines with opposite orientations. This is 
easier to achieve for the striped wind geometry during the
collision processes, but is more difficult for the helical geometry.
In the majority of the locations for the helical configuration, 
the field lines in the two colliding shells have the same 
orientation. However, astrophysical systems are not perfectly
symmetric systems. One may imagine some scenarios to trigger
reconnection events at some locations. For example,
current-driven kink instability may develop in the jet
\citep[e.g.][]{mizuno09b}, which would introduce a slight 
misalignment of the magnetic field axes in two consecutive 
``shells''. This would result in a small cross section near the 
magnetic axes that have opposite orientations in the two shells 
(Fig.\ref{fig:trigger}). The first seed reconnection would 
then occur. 

\item Once reconnection seed event is triggered, the rapid 
energy ejection from the reconnection layer would disturb the
nearby ambient plasma and field lines to make the region 
more turbulent.
The reconnection layer is also subject to various instabilities, 
e.g., the two-stream instability that develops as a consequence 
of reconnection acceleration. The turbulence stirs the plasma
to allow more opposite-orientation field lines approach each
other, so that further reconnection events occur. 
These new reconnection events would
eject more energy to make the ejecta more turbulent, leading
to a reconnection-turbulence avalanche. This would result in a 
runaway catastrophic release of the stored magnetic field energy. 
Particles are accelerated either directly in the reconnection 
regions, or stochastically 
in the turbulent regions, which radiate synchrotron
photons that power the observed GRB emission. The discharge
process ceases when $\sigma$ is reduced to $\sigma_{end} \leq 1
\ll \sigma_{ini}$. This is one ICMART event, which would compose
one fundamental unit of a GRB (one pulse). Other collisions
that trigger other reconnection-turbulence avalanches would
give rise to other pulses in the GRB.
A GRB event is the superposition of many individual ICMART
events. 
\end{itemize}

GRB events are relativistic in nature. Although the condition
(\ref{condition}) or (\ref{condition1}) gives the requirement
for powering a GRB with global turbulent reconnection. In order
to achieve such a relativistic reconnection/turbulence avalanche,
some physical conditions may be needed.

The first relevant condition is the strength of the collisions
that distort the field line configurations. We first comment on 
the role of internal shocks in these collisions.
A common misconception is that shocks do not exist in the high-$\sigma$ 
regime. This is not always true. It is relevant for the 
interaction between a strongly magnetized (high-$\sigma$) shell 
and a non-magnetized ($\sigma=0$) medium (e.g. deceleration of a
magnetized shell by the ISM). No shock can form 
in the high-$\sigma$ shell if the thermal pressure in the 
shocked non-magnetized medium is smaller than the magnetic pressure 
in the magnetized shell 
\citep{zhangkobayashi05,mizuno09}\footnote{The reverse shock
forming condition introduced by \cite{giannios08b} is erroneous. 
These authors claimed that there should be no reverse shock when 
$\sigma > 0.3$, but the later numerical simulations performed by 
the same group suggests that a weak reverse shock does exist in
the $\sigma > 0.3$ regime \citep{mimica09}. The numerical result 
is consistent with \cite{zhangkobayashi05} and \cite{mizuno09}
who suggested that the only physical condition to define the 
existence of a reverse shock in a magnetized ejecta is whether
the forward shock pressure exceeds the magnetic pressure of the
ejecta.}. 
On the other hand, for the collision between two shells with the 
same $\sigma$ value, even a moderate collision
would give rise to an excess pressure (ram pressure) that exceeds
the magnetic pressure. This would lead to
a pair of shocks passing through both shells. 
For a given $\sigma$ value, the strength of the shock increases 
with $\Gamma_{fs}$, and can reach more than 50\% of the strength
of a $\sigma=0$ shock once $\Gamma_{fs}$ exceeds 3 
\cite[Figs 1 \& 2 of][]{zhangkobayashi05}. So in the ICMART model 
essentially all the collisions are accompanied by 
internal shocks, but the main energy dissipation is not 
through internal shocks. What the shocks do is to disturb the velocity,
density and magnetic fields. These distortions are reinforced
via turbulence, which reduce $\lambda_\|$, 
until an ICMART event is eventually triggered. Without numerical
simulations, it is hard to quantitatively analyze the role of
internal shocks in the high-$\sigma$ regime. Qualitatively, for
the same $\sigma$ value, a larger
relative Lorentz factor between the two shells ($\Gamma_{fs}$)
tends to introduce a larger perturbation in the ordered magnetic
fields. Within the
rest frame of shell 1, the ram pressure exerted by 
shell 2 is $P_{ram,21} = \Gamma_{21}^2 \rho'_2 c^2$, which would
be larger than its own magnetic pressure $P_{B,1} = {B'_1}^2/8\pi
= (1/2) \rho'_1 c^2 \sigma$, when
\begin{equation}
\Gamma_{21} \geq \left(\frac{1}{2} \sigma \frac{\rho'_1}{\rho'_2}
\right)^{1/2}
\label{shock-perturbation}
\end{equation}
is satisfied. This is the condition for a strong 
shock-induced-perturbation to occur.

A second relevant condition is that the outflow from a reconnection
event has to be relativistic. This would induce relativistic turbulence,
which is essential to induce turbulence-related GRB temporal
variability (see more in Sect.\ref{sec:variability} below).
The relativistic extension of Sweet-Parker and Petschek scenarios 
have been discussed by \cite{blackman94} and \cite{lyutikov03c}.
Based on energy conservation conditions, these authors argued that
under certain conditions, both the inflow and outflow speeds ($V_{in}$ 
and $V_{out}$) can reach the
relativistic regime. These treatments did not consider force balance
across the reconnection layer, an effect becoming very important
in the relativistic regime. \cite{lyubarsky05} evaluated this 
effect and correctly concluded that $V_{in}$ can never achieve the
relativistic speed. In any case, in long term 
the condition imposed by energy conservation should be satisfied, 
so that the outflow can achieve a relativistic speed under certain 
conditions \citep[][M. Lyutikov, 2010, private communication]{lyubarsky05}.
This condition is essential to induce relativistic turbulence.
The condition for a relativistic outflow may be approximated
as $s < \sigma$ \citep[e.g.][]{lyutikov03c}, or, in terms of the
Bohm diffusion limit
\begin{equation}
\lambda_\| < \sigma r'_{B,e} \simeq 9.3\times 10^3 ~{\rm cm}~
\sigma_1 \gamma_{e,3} L_{w,52}^{-1/2} \Gamma_{2.5} R_{15}
\left(\frac{1+\sigma}{\sigma}\right)^{1/2}~.
\label{condition2}
\end{equation}
This condition is more demanding than Eq.(\ref{condition1}), 
which means that once relativistic turbulence is excited,
the turbulent reconnection avalanche can naturally power
a GRB.

\subsection{Preference of a large emission radius}
\label{sec:largedistance}

The ICMART events occur preferably at large radii from the
central engine. This is because triggering
an ICMART event requires that the conditions Eq.(\ref{condition})
and probably also Eq.(\ref{condition2}) are satisfied, which
demands $\lambda_\|$ being small enough. Early
collisions (typically at small radii) only serve to distort the
field lines and do not directly trigger ICMART events. 
Statistically, shells that collide at large radii tend to have
more distorted field configurations and hence, can more easily
trigger ICMART events (Fig.\ref{fig:cartoon}). 

Observationally, it is not easy to infer the GRB emission radius
$R_{\rm GRB}$ from the MeV data alone. However, by combining 
radiation information from other wavelengths, there are three
independent, indirect ways to constrain $R_{\rm GRB}$. 
(1) Swift observations suggest that the early X-ray afterglow 
light curves are dominated by a steep decay component 
\citep{tagliaferri05}, which is best interpreted as the
the high-latitude emission of the prompt GRB as the emission
is abruptly ceased \citep{zhang06}. Within such an interpretation,
the duration of the tail defines the minimum of $R_{\rm GRB} 
\theta_j$. The data suggest that $R_{\rm GRB}$ is typically
large (say, $R_{\rm GRB} \geq 10^{15}$ cm) if the high-latitude
emission interpretation is valid \citep{lyutikov06,kumar07}.
(2) Some GRBs have prompt optical detections or upper limits. 
If the optical emission arises from the same region as the 
MeV emission, then the brightness of the optical emission can
be used to derive a constraint on $R_{\rm GRB}$, based on the
argument that synchrotron self-absorption should not suppress 
the optical flux. The bright optical prompt emission of 
GRB 080319B requires that the optical emission region has to
be greater than $10^{16}$ cm \citep{racusin08,kumarpanaitescu08},
so that $R_{\rm GRB} > 10^{16}$ cm is inferred if the MeV emission 
and optical emission originate from the same region 
\citep[c.f.][]{fan09,zou09}. A systematic analysis of other
prompt optical data \citep{shen09} suggests that data are
consistent with $R_{\rm GRB} \geq 10^{14}$ cm. 
(3) If GeV emission is from the same region as the MeV component,
as is inferred from most Fermi/LAT GRBs \citep{zhang10}, 
the detection of the high energy photons can be used to
constrain both the emission radius and the bulk Lorentz factor
\citep{gupta08}. Applying this method to GRB 080916C, the derived
$R_{\rm GRB}$ is typically greater than $10^{15}$ cm 
\citep{zhangpeer09}. We can see that the three independent
pieces of information generally point towards a consistent
picture, namely, the GRB emission radius $R_{\rm GRB}$
is relatively large. This is consistent
with the expectation of the ICMART model.

\section{Merits and implications of the ICMART model}
\label{sec:merits}

The ICMART model invokes internal collisions as the trigger of
magnetic reconnection/turbulence avalanche. It therefore carries
the merits of the internal shock model, namely, the emission
site is ``internal'', and the central engine activity defines
the main variability time scale (the pulses in the 
light curve)\footnote{Again this is also the merit of the
dissipative photosphere model which does not demand internal
collisions.}. 
On the other hand, it can overcome or alleviate some of
the drawbacks of the IS model
as discussed in \S2.1, which we elaborate in the following.

\subsection{Radiative efficiency}

Let's assume that two shells [$((m_1, \Gamma_1)$ and 
$(m_2, \Gamma_2)$] with an initial $\sigma_{ini}$ 
collide with each other, and that the collision triggers a 
reconnection/turbulence cascade that results in a catastrophic
discharge of the magnetic energy. This leads to conversion of 
a significant amount of magnetic energy to internal 
energy of the fluid, and hence, to radiation. Let's envisage
a picture where the two shells merge with a much lower magnetization 
parameter $\sigma_{end}$ by the end of such an inelastic collision.
Energy conservation 
\begin{equation}
(\Gamma_2 m_2 + \Gamma_1 m_1)(1+\sigma_{ini})
=\Gamma_m (m_1+m_2+U') (1+\sigma_{end}) 
\end{equation}
and momentum conservation
\begin{equation}
(\Gamma_2 \beta_2 m_2 + \Gamma_1 \beta_1 m_1)(1+\sigma_{ini})
=\Gamma_m \beta_m (m_1+m_2+U') (1+\sigma_{end})
\end{equation} 
give the same
solution for $\Gamma_m$ [Eq.(\ref{Gammam})] as the IS model.
The energy dissipation efficiency, on the other hand, is
much larger, i.e.
\begin{eqnarray}
\eta_{_{\rm ICMART}} & = & \frac{\Gamma_m U'}{(\Gamma_1 m_1c^2+\Gamma_2 m_2c^2)(1+\sigma_{ini})}
\nonumber \\
& = & \frac{1}{1+\sigma_{end}} - \frac{\Gamma_m(m_1+m_2)}
{(\Gamma_1 m_1 + \Gamma_2 m_2) (1+\sigma_{ini})} \nonumber \\
& \simeq & \frac{1}{1+\sigma_{end}}~~~~~({\rm if}~ \sigma_{ini} \gg 1).
\label{eta}
\end{eqnarray}
This is $\sim 50\%$ if $\sigma_{end} \sim 1$, and can reach
$90\%$ if $\sigma_{end} \sim 0.1$. This can naturally account
for the observed high radiative efficiency of most GRBs \citep{zhang07a}.

\subsection{Particle acceleration and the fast cooling problem}

In an ICMART event, particles can be accelerated in the following
two ways. First, particles are accelerated directly in the reconnection 
zone. This happens when particles are bounced back and forth between the 
oppositely directed approaching magnetic fluxes and gain energy during 
each gyration. It is similar to shock acceleration and is an efficient 
1st order Fermi process \citep{dalpino05,lazarian09}.
The second acceleration mechanism is the
stochastic acceleration of particles in the turbulent region, 
which is a 2nd order Fermi process. 

We defer the detailed discussion on particle acceleration and radiation 
spectrum of the ICMART model to a future work, but qualitatively discuss 
the difference between this model and the shock model. Within the 
standard internal shock model, particles are accelerated at the
shock front via the 1st-order Fermi mechanism
\citep{spitkovsky08,baring09}. After the shock passes through the shell,
the downstream hot electrons would cool rapidly through synchrotron
and inverse Compton emission. Once an electron is accelerated to
a high energy, there is no further heating process, and the electron
suffers radiative cooling. Such a scenario inevitably introduces
the so-called fast cooling problem as discussed in
\S\ref{sec:fastcooling}, unless a much small-scale magnetic
field is introduced \citep{peerzhang06,asano09}.

In the ICMART scenario, electrons do not undergo such a ``one-shot''
acceleration like in the shock scenario. Rather, they would be
accelerated repetitively. This is because the ICMART avalanche happens
globally in the dissipation region (unlike just near the shock front
in the shock scenario). There is no distinct separation of the particle
acceleration region and the cooling region. As a result, any particle
would undergo multiple accelerations in the reconnection regions
either directly or stochastic scattering of the constantly evolving 
turbulent field. For the 1st-order reconnection acceleration
scenario, particle acceleration time could be much shorter than the 
cooling time scale, so that a power-law distributed electron
population can be accelerated up to an energy at which cooling
time scale balances the acceleration time scale (see 
Sect.\ref{sec:Ep} below). This is similar
to the shock scenario, in which the fast cooling problem remains. 
On the other hand, for the 2nd-order turbulence acceleration 
scenario the acceleration time can be long so that one may achieve 
a balance between cooling and heating.
In the past, this was referred to as the ``slow volume heating'' 
scenario, which invokes Compton upscattering off emission from 
a dissipative photosphere 
\citep[e.g.][]{thompson94,ghisellini99,stern04,peer06,giannios07,beloborodov09}.
The electron energy of these previous slow heating models is 
typically sub-relativistic, say, 10s-100s keV. 
In the ICMART scenario discussed here, the
slow heating due to turbulent acceleration would happen in the 
relativistic regime if a significant fraction of the magnetic
energy is transferred to particle energy via the 2nd
order stochastic turbulence acceleration. This is due to the
combination of two effects: a much reduced electron number in
a high-$\sigma$ flow (reduced by a factor $1+\sigma$) and a much 
lower magnetic field at a much larger emission radius than
photosphere (i.e. $R_{_{\rm ICMART}} \gg R_{ph}$). 
To show this, we assume that a 
fraction $\xi$ of comoving magnetic energy is consumed through
slow heating by turbulence acceleration. The average heating 
rate per electron in the emission region is $\dot E'_{heat} 
\simeq \xi U'_B \varepsilon_e/ \Delta t' n'_e$, where $\varepsilon_e$
is the fraction of the dissipated energy that is distributed to
electrons. Noticing that the dominant cooling mechanism in a
high-$\sigma$ flow is synchrotron radiation, the average cooling
rate can be estimated as $\dot E'_{cool} \simeq (4/3) \gamma_e^2
\sigma_{\rm T} c U'_B$. Balancing heating and cooling to make
$\dot E'_{heat}=\dot E'_{cool}$, one can solve for the critical
electron energy below which slow heating dominates, i.e.
\begin{eqnarray}
\gamma_{e,h} & = & \left(\frac{3 \xi \varepsilon_e}{4\sigma_T c \Delta t'
n'_e}\right)^{1/2}  \simeq 2.6\times 10^3 \left(\frac{\xi \varepsilon_e}
{Y}\right)^{1/2} \nonumber \\
& \times &  L_{w,52}^{-1/2} \Gamma_{2.5}^{1/2} R_{15} m^{1/2} 
(1+\sigma)_1^{1/2} \Delta t_{-3}^{-1/2}~,
\label{gammaeh}
\end{eqnarray}
where $\Delta t'=\Gamma \Delta t$ has been used, and $\Delta t$ has
been normalized to $10^{-3}$ s, the minimum variability time scale
detected in GRBs. Equation (\ref{gammaeh}) therefore gives the 
maximum electron Lorentz factor below which slow heating dominates
and fast cooling problem is no longer relevant. The corresponding 
synchrotron photon energy is then
\begin{eqnarray}
E_h & \sim & \hbar\Gamma \gamma_{e,h}^2 \frac{eB'}{m_e c}(1+z)^{-1}
\sim 23~{\rm keV} \left(\frac{\xi\varepsilon_e}{Y}\right) 
\nonumber \\
& \times & 
L_{w,52}^{-1/2}\Gamma_{2.5} R_{15} m (1+\sigma)_1 
\left(\frac{\sigma}{1+\sigma}
\right)^{1/2} \Delta t_{-3}^{-1} \left(\frac{1+z}{2}\right)^{-1}~.
\end{eqnarray}

In reality, both acceleration mechanisms would operate in the
ICMART regions. The 1st-order reconnection acceleration likely
dominates, which would define $E_p$ and high energy emission of
the GRB. If the 2nd-order turbulence acceleration can transfer
a good fraction energy to particles so that 
$\xi$ is not $\ll 1$, then $E_h$ is close to $E_p$
(Eq.[\ref{Epsyn-ICMRT}] discussed in the next section). 
The fast cooling problem may be alleviated.

\subsection{Electron number and $E_p$}
\label{sec:Ep}

The electron number excess problem of the internal shock model 
(\S\ref{sec:enumber}) is naturally overcome in the ICMART model.
This is because the baryon-associated electron number is simply 
$(1+\sigma)^{-1}$ of that in the baryon-dominated model for the 
same $L_w$. In principle, pair production can increase the lepton 
number, but the large emission site related to the ICMART model
is way above the pair photosphere 
\citep{meszaros02b,kobayashi02,peer06} so that
pair production may not be effective.
This makes the Lorentz factor of each electron increase by a factor
of $(1+\sigma)$ with respect to the baryon-dominated case, naturally
raising $E_p$ to the $\gamma$-ray band.

To quantitatively address this point, we investigate the typical 
energy of synchrotron radiation in a high-$\sigma$ flow. Synchrotron
radiation is the dominant mechanism in a Poynting-flux-dominated flow,
since the photon energy density (relevant for inverse Compton
scattering) is much smaller than the magnetic energy density.
The spectral peak energy $E_p$ in the ICMRT model can also be
expressed as Eq.(\ref{Epsyn}). The comoving magnetic field 
can be estimated according to Eq.(\ref{B'}).
The factor $\sigma/(1+\sigma)$ ranges from $\sim 1$ (for $\sigma_{ini}
\gg 1$ in the beginning of an ICMART event) to $\sim 1/2$ (for
$\sigma_{end} \sim 1$ at the end of the ICMART event).
Assuming that the particle acceleration is dominated by the
1st order reconnection acceleration, and that the energy
release from magnetic dissipation is distributed to
electrons and protons in the fractions of $\varepsilon_e$ and
$\varepsilon_p$ with $\varepsilon_e+\varepsilon_p=1$ (to be 
differentiated from $\epsilon_e$, $\epsilon_p$ and $\epsilon_B$
in the shock case), one can write
\begin{equation}
\frac{L_w \eta}{4\pi R^2 c \Gamma^2}=n'_p \bar\gamma_p m_p c^2
+n'_e \bar\gamma_e m_e c^2 = n'_e \bar\gamma_e m_e c^2 / \varepsilon_e.
\end{equation}
The comoving electron number density can be calculated by
Eqs.(\ref{ne'}) and (\ref{np'}). This gives
\begin{equation}
\bar\gamma_e=\eta\varepsilon_e (1+\sigma)\frac{m_p+Y m_e}{Y m_e}~.
\end{equation}
The minimum electron Lorentz factor can be derived by 
$\gamma_{e,m}=\phi(p) \bar\gamma_e$. In the ICMART model,
the observed gamma-ray luminosity can be expressed as
\begin{equation}
L_\gamma=L_w \eta \varepsilon_e~.
\end{equation}
one can then write $E_p$ as
\begin{eqnarray}
E_{\rm p,ICMART} & \sim & \frac{m_e c^2}{B_q} [\phi(p)]^2
\left(\frac{m_p+Y m_e}{Y m_e}\right)^2 
\left(\frac{L_\gamma}{R_{_{\rm ICMART}}^2 c}\right)^{1/2} \nonumber \\
& & \times (1+z)^{-1} \left[
\left(\frac{\sigma}{\eta\varepsilon_e(1+\sigma)}\right)^{1/2}
(\eta\varepsilon_e)^2(1+\sigma)^2 
\right] \nonumber \\
& \simeq & 160~{\rm keV}~ \left(\frac{\phi(p)}{1/6}\right)
L_{\gamma,52}^{1/2} R_{_{\rm ICMART,15}}^{-1}
\nonumber \\
& \times & (\eta \varepsilon_e)^{3/2} \sigma_{1.5}^{2}
\left(\frac{1+z}{2}\right)^{-1}~.
\label{Epsyn-ICMRT}
\end{eqnarray}
Compared with the $E_p$ expression in the IS model
(Eq.[\ref{Epsyn-IS}]), the ICMART model can easily make 
$E_p$ in the range of several hundreds of keV (as observed) by 
requiring $\sigma \sim$ several 10s. 

By reducing the number of electrons, one can also naturally
overcome the self-absorption problem introduced by the
electron number excess problem \citep{shen09}.

\subsection{Amati/Yonetoku relation}

Comparing Eq.(\ref{Epsyn-IS}) and Eq.(\ref{Epsyn-ICMRT}), we can
see that  $E_p$ of the IS model and ICMART model both have
the dependence $E_p \propto L_{\gamma}^{1/2}$. The problem
of the IS model (as discussed in \S\ref{sec:amati}) is that
$R_{\rm IS}$ has an additional dependence on $\Gamma$, which is 
correlated to $L_\gamma$ \citep{liang10}. This completely
destroys the $E_p \propto L_{\gamma}^{1/2}$ dependence.
In the ICMART model, $E_p$ has the dependence of $\propto L^{1/2}
\sigma^2 R_{_{\rm ICMART}}^{-1}$. It is hard to analytically
investigate how $\sigma$ and $R_{_{\rm ICMART}}$ influence
the apparent $E_p \propto L_{\gamma}^{1/2}$ dependence. The
sensitive dependence on $\sigma$ suggests that a small variation
of $\sigma$ would lead to a significant variation of $E_p$. On
the other hand, the shallow dependence of $\sigma$ on $\sigma_0$
(2/3 power) and $R$ (-1/3 power) suggests that the variation of
$\sigma$ may not be significant. Also, an outflow with a higher 
$\sigma$ would be more difficult to trigger an ICMART event,
since it takes more stringent criteria (e.g. a larger $\Gamma$
contract and more collisions) to distort the field lines (e.g. 
Eq.[\ref{shock-perturbation}]). Therefore an ICMART event with
a higher $\sigma$ tends to happen at a larger $R_{_{\rm ICMART}}$.
This would partially compensate the $E_p$ scatter due to $\sigma$
variation. If the $\sigma^2  R_{_{\rm ICMART}}^{-1}$ factor has a 
weak dependence on $L_{\gamma}$ and does not have a large scatter, 
then one may obtain the Amati/Yonetoku correlation. Numerical 
simulations of various processes (e.g. magnetic field 
distortion during collisions, reconnection physics, and 
the trigger condition of ICMART events) are needed to
verify or disproof such a speculation.

\subsection{Weak photosphere}

Within the ICMART picture, the $\sigma$ value remains moderately
high at $R_{\rm GRB}$ before strong magnetic field dissipation
happens. This requires that the magnetic
acceleration and dissipation effects are not prominent at the
smaller radii. This guarantees a weak photosphere emission component
(energy smaller by a factor of $(1+\sigma)^{-1}$ with respect
to the baryon-dominated model), which satisfies the constraints of
the GRB 080916C data \citep{zhangpeer09}. 

\section{Predictions of the ICMART model}
\label{sec:predictions}

The ICMART model makes a list of  predictions that may be used
to differentiate it from other prompt emission models using the 
observational data.

\subsection{Variability time scales: engine and turbulence components}
\label{sec:variability}

The most important prediction of the ICMART model is that there
are two mechanisms that define the observed GRB variability
time scales. Since the emission is triggered by internal
collisions, the internal wind irregularity inevitably leaves an
imprint on the light curve. This engine-defined variability
is similar to that of the internal shock model, and is 
relevant to a broader (``slow'') component of variability.
The characteristic time scale of this component is defined by the
angular spreading time at the large emission radius, which reads
$R_{_{\rm ICMART}}/\Gamma^2 c \sim 0.3~{\rm s}~R_{_{\rm ICMART,15}} 
\Gamma_{2.5}^{-2}$. Phenomenologically, this corresponds to
individual broad ``pulses'' of the light curves, which is visible in 
some GRBs. On the other hand, within 
each ICMART emission region (the broad pulse), there are
many turbulent regions from which highly variable emission is 
released. This corresponds to a second variability component
with very small time scales. In the previous
GRB models, the variability time in all scales are either
defined by the central engine \citep{sari97} or by relativistic
turbulence \citep{narayan09}. The ICMART model suggests
that the observed GRB variability is the superposition of these 
two components. The two variability components 
may be differentiated via a proper temporal analysis of GRB
light curves. 

\begin{figure}
\plotone{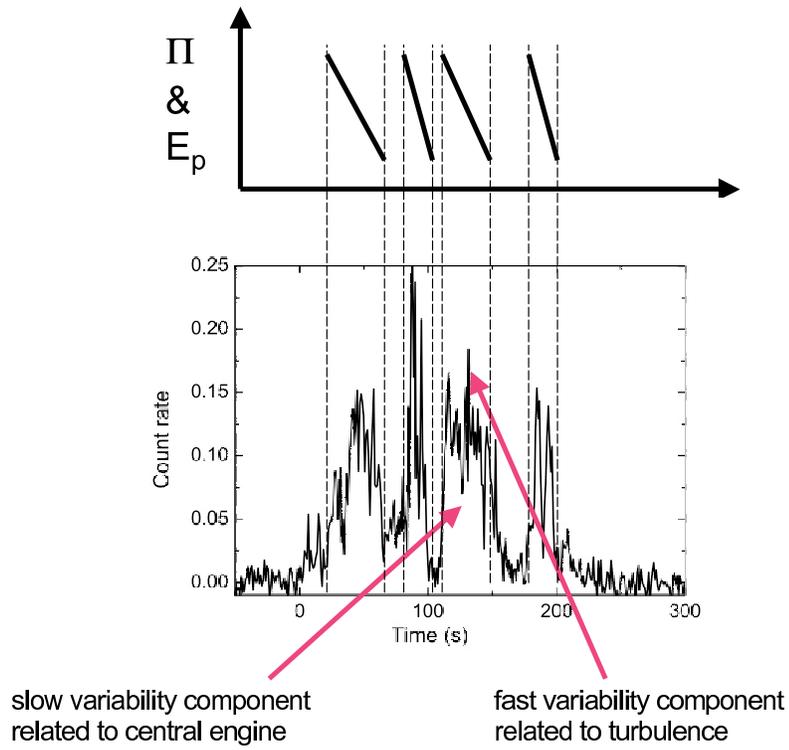}
\caption{An example of GRB light curve that shows two variability time scales. The light curve of 
GRB 050117 is extracted from the UNLV GRB
group website http://grb.physics.unlv.edu/$\sim$xrt/xrtweb/050117/050117.html.
The predictions of decreasing gamma-ray polarization degree $\Pi$ and the spectral
peak energy $E_p$ within individual pulses are indicatively presented. Detailed decaying functions
would be different depending on the details of evolution of magnetic field configuration, $\sigma$
value, as well as balance between heating and cooling of electrons. The general decreasing trend
is robust.
}
\label{fig:lightcurve}
\end{figure}

Visually some GRB light curves do show several visible slow bumps 
superposed by rapid variability features (Fig.\ref{fig:lightcurve}). Power 
density spectrum analyses using Fourier transform \citep{beloborodov00b},
however, did not reveal distinct two-component variability
distributions. Nonetheless, studies using some more sophisticated
statistical methods have started to reveal the evidence for two 
variability components at least for some GRBs 
(e.g. Shen \& Song 2003; Vetere et al. 2006,
see also H. Gao et al. 2010, 
in preparation; R. Margutti et al. 2010, in preparation).

\subsection{$E_p$ evolution within a pulse}

In the ICMART model, one ICMART event (triggered by one 
collision) corresponds to one pulse in the GRB light curve. 
During each event, since the magnetic energy is continuously
converted to the particle energy and then released as 
radiation, one may approximately treat the plasma
as having $\sigma$ continuously decrease with time 
while the lepton number not changed. According to
Eq.[\ref{Epsyn-ICMRT}]), the peak energy in the ICMART model
evolves as $E_p \propto L_\gamma^{1/2} \sigma^2$. Although
$L_\gamma$ increases shallowly during the rising phase, the
steep dependence on $\sigma$ compensates this. Noticing that
during the rising phase $L_\gamma$ only increases by a factor
of a few while $\sigma$ drops 1-2 orders of magnitudes, one
expects that in general $E_p$ decreases throughout a pulse
(the fundamental radiation unit) in GRB prompt 
emission (Fig.\ref{fig:lightcurve}), probably with a 
steepening in decay after the light curve peak.
Observationally, the majority 
of GRB pulses indeed show a hard-to-soft evolution behavior
\citep{norris86,ford95,lu10}, which is consistent with the 
expectation of the ICMART model. 
On the other hand, a small fraction of pulses show the 
so-called ``tracking'' behavior, i.e. the hardness of the 
spectrum is positively correlated with the flux variation 
across the pulse \citep{golenetskii83}. These pulses cannot 
be related to an ICMART event. It may, however, be interpreted
by emission from a turbulent eddy, with the rising and decaying
components corresponding to entry and exit of the eddy from the
field of view ($\Gamma^{-1}$ cone). 
This requires the existence of a large scale
eddy that can give rise to a broad pulse in the GRB light curve.

\subsection{Gamma-ray polarization evolution within a pulse}

Since the ICMART process destroys the ordered global
magnetic fields in the ejecta, one naturally expects an
evolution of the linear polarization degree in the 
{\em $\gamma$-ray} emission during each ICMART unit, i.e.
a pulse. The initial gamma-ray polarization degree may be
close to the maximum value achievable for synchrotron 
emission in an ordered magnetic field, i.e. $\Pi\sim 50\%-60\%$
\citep{lyutikov03b,granot03}. At the end of the ICMART event,
the ordered field structure would be largely destroyed. 
However, even in the fully turbulent limit, the net polarization
degree would not become zero, since in the strong field regime,
turbulence is anisotropic with eddies elongated along the 
local field lines. The average polarization
in the $\Gamma^{-1}$-cone field of view would not be zero. Without
detailed numerical simulations of an ICMART event (which requires
a dissipative, relativistic MHD with the details of reconnection
processes delineated, and which is beyond the scope of all the current 
GRB motivated MHD simulations), it is hard to quantitatively predict
the polarization degree at the end of the ICMART event. 
Nonetheless, a value of $\Pi \sim $ a few$\% - 10\%$ may be 
reasonable. Since different pulses correspond to different
ICMART events in our model, we expect a significant polarization
degree evolution during the GRB prompt phase, with $\Pi$ starting
with a high value at the beginning of each pulse and evolving
from high to low across the pulse (Fig.\ref{fig:lightcurve}). 
For bursts with overlapping 
pulses, the $\Pi$ evolution would be more complicated since it
reflects the superposition of the contributions of different 
pulses.

Observationally, there is no robust gamma-ray polarization 
detection yet. \cite{coburn03} reported $\Pi=80\%\pm20\%$
for GRB 021206 using the RHESSI data. However the conclusion
cannot be confirmed by an independent analysis of the same
data \citep{rutledge04,wigger04}. \cite{willis05} derived
lower limits of the polarization degree $\Pi>35\%$ and $\Pi>50\%$
for GRB 930131 and GRB 960924, respectively, using the BATSE
Albedo Polarimetry System (BAPS), but could not strongly constrain
the degree of polarization beyond a systematics-based estimation.
Another polarization measurement for the INTEGRAL burst 
GRB 041219A \citep{kalemci07,mcglynn07} is also subject to
large uncertainty. Nonetheless, these studies show the
tentative evidence that GRB $\gamma$-ray emission may be
polarized. A true breakthrough could be made by one of several
proposed $\gamma$-ray polarimeters that are suitable to detect
polarized $\gamma$-rays from transient sources, e.g. POET
\citep{hill08}, POLAR \citep{lamanna08}, and PoGO \citep{mizuno05}.
For example, for completely ordered magnetic fields, POET can 
measure the polarizations of $>30\%$ of bursts detected by 
the mission \citep{toma09b}.
The unique time-dependent $\Pi$ evolution pattern predicted for
the ICMART model may be tested by future observations with these
detectors.

\subsection{Polarized external reverse shock emission}

Regardless of the composition of the GRB ejecta, the outflow
is eventually decelerated by a circumburst medium, be it a
constant density interstellar medium (ISM), or a stratified 
stellar wind \citep{meszarosrees97,sari98,dailu98c,chevalier00}. 
Before entering the self-similar deceleration regime
\citep{blandford76}, a reverse shock may propagate into the
ejecta to decelerate it. The composition of the ejecta does
affect the radiation signature from the reverse shock
\citep{zhangkobayashi05}. In particular, the reverse shock
emission is expected not to be bright or is completely suppressed
if $\sigma$ is very small \citep{fan04,nakarpiran04,jinfan07} 
or very large \citep{zhangkobayashi05,mimica09}. A bright
optical flash may be detected if $\sigma \lesssim 1$, as is 
discovered in some GRBs such as GRB 990123
\citep{fan02,zhang03,kumar03,gomboc08}.
At a slightly higher $\sigma \geq 1$, the reverse shock 
emission brightness drops significantly.

In the ICMART model, the final $\sigma$ value after the magnetic
dissipation would be close to (maybe slightly below)
unity, based on an energy equipartition argument. The
exact $\sigma$ value is hard to predict and is subject to many
uncertainties. In any case, similar to the discussion of prompt
$\gamma$-ray polarization above, the reverse shock optical emission
should also be somewhat polarized (say, a few percent). If 
polarimetric observations can be carried out during the early rapidly 
decaying phase due to the reverse shock emission
(say, in the optical band), a polarized 
optical signal should be detected. Observations of early optical
polarimetry are now regularly carried out by some groups,
e.g. the Liverpool optical polarization
observation group. Measurements of early optical polarimetry 
have been made for two GRBs. For GRB 060418, which shows a smooth
deceleration bump that is likely due to the forward shock emission,
an upper limit of $\Pi < 8\%$ was set up \citep{mundell07}. Since
the forward shock is the emission of shocked circumburst medium,
this low level polarization is entirely consistent with the
theoretical expectation. On the other hand, recently the team
detected a $\Pi=10\% \pm 2\%$ optical polarization from GRB 090102
\citep{steele09}. The polarized signal was measured during the 
rapid decay phase of the early optical light curve, which is
consistent with being the reverse shock emission with $\sigma < 1$.
This observational fact is consistent with the
expectation of the ICMART model. More future early optical
polarimetric data are needed to further test this
prediction of the ICMART model.

\subsection{No detectable synchrotron self-Compton spectral component
during GRB prompt emission}

According to the ICMART model, throughout the GRB prompt emission phase,
one has $\sigma \geq 1$ in the emission region. The Compton parameter
${\cal Y}$ is therefore $<<1$. The magnetic field energy density is
much larger than the synchrotron photon energy density, so the SSC 
process is greatly suppressed. One therefore expects no extra high energy
emission features other than extrapolation of the MeV spectrum (Band
spectrum) to high energy. This is consistent with the observations of
the majority of Fermi LAT GRBs, including GRB 080916C \citep{zhang10}.

\section{GRB 080916C}
\label{sec:GRB080916C}

In this section we discuss GRB 080916C within the ICMART model.
This is the first bright Fermi/LAT burst whose time-dependent 
broad band spectra are measured. 
The observational results are somewhat surprising for
theorists. Three theoretically motivated features, namely the
pair cutoff feature at high energy, the SSC
component at high energy, and the quasi-blackbody photosphere
emission component are all missing. Instead, the time dependent
spectra are a set of nearly featureless ``Band'' functions ranging
from $\sim 10$ keV to $\sim 10$ GeV \citep{abdo09,zhang10}.
In order to interpret this burst within the baryonic models,
one needs to introduce several spectral components that conspire
to mimic a Band function covering 6-7 orders of magnitude.
For example, the model developed by \cite{toma09} invokes
SSC in the internal shock, upscattered cocoon emission, as well
as another synchrotron emission component to model the
observed Band function spectrum. The neutron-heating photosphere
model developed by \cite{beloborodov09} needs to attribute the
very high energy emission to a different component \citep[e.g. 
from the external shock,][]{kumar09,ghisellini09}. The low energy
slope below $E_p$ is predicted to be $\alpha=0.4$, much harder
than the observed $\alpha \sim -1$ slope in most of the epochs
\citep{abdo09}. 
Another model developed by \cite{li10} applies the standard 
internal shock synchrotron emission to account for the emission 
near $E_p$, while invokes a series of residual internal shocks from
which the IC emission spectra are superposed to account for the
emission above 100 MeV. Finally,
\cite{razzaque09} invokes a standard leptonic synchrotron 
component to interpret the MeV emission, while introduces a 
hadronic proton synchrotron emission component to account for 
emission above 100 MeV. All these models require fine tuning of
their model parameters to account for a very simple broad-band 
Band spectrum that covers 6-7 orders of magnitude in all 
the time intervals, as is revealed by detailed time-dependent 
spectral analyses \citep{abdo09,zhang10}.
Furthermore, none of these models
have calculated the contribution of the photosphere emission
(except Beloborodov 2010, who discussed the photosphere model
itself), but rather raised some simple arguments (without
calculation) to avoid the missing bright photosphere problem raised 
by \cite{zhangpeer09}. A careful calculation by \cite{fan10} showed 
that the Zhang-Pe'er argument cannot be circumvented even under 
the most favorable condition to ``hide'' the photosphere component.

The featureless Band function can be straightforwardly interpreted
within the ICMART model. The lack of the three theoretically
expected features is understandable within the ICMART model.
1. Since the emission radius is large for ICMART events 
(\S\ref{sec:largedistance}), the pair production opacity is
reduced, so that the rest-frame $\sim 70.6$ GeV photon can escape 
the emission region.
One therefore does not expect a pair cutoff feature in the 
spectrum. 2. Since the outflow is magnetically dominated in the
comoving frame (i.e. $\sigma > 1$), the magnetic energy density
is naturally much larger than the synchrotron photon energy
density. The Compton parameter ${\cal Y}$ is $\ll 1$. One therefore
does not expect a SSC component in the high energy regime.
3. The ``hot'' component only carries a luminosity which is
$(1+\sigma)^{-1}$ times $L_w$. As a result, the photosphere
emission is much dimmer, which can be hidden beneath the 
non-thermal emission component from the ICMART event
\citep{zhangpeer09}.

Within the ICMART model, the entire Band function spectrum
is one emission component, which is powered by synchrotron
emission of the electrons accelerated in the ICMART events.
GRB 080916C has an observed $E_p$ ranging from $400 - 1200$ 
keV in various time intervals \citep{abdo09}. At $z\sim 4.35$, 
its average luminosity in the first two main pulses is of the 
order of $\sim (1-2) \times 10^{53}~{\rm erg~s^{-1}}$. 
From Eq.(\ref{Epsyn-ICMRT}), one could estimate the required
$\sigma$ for this burst
\begin{equation}
\sigma \sim (38-67) (\eta \varepsilon_e)^{-3/4} R_{_{\rm ICMART,15}}^{1/2}~.
\label{080916Csigma}
\end{equation}
Since $(\eta \varepsilon) \leq 1$, the estimate 
Eq.(\ref{080916Csigma}) is fully consistent with the 
independent constraint $\sigma > (15-20)$ based on the
non-detection of the photosphere component \citep{zhangpeer09}.
One can also estimate the maximum synchrotron emission energy
in the ICMART model. Similar to the shock model, the maximum
electron Lorentz factor is defined by equating the comoving
acceleration time scale $t'_{acc} = \kappa \gamma_e m_e c/eB'$
(where $\kappa \geq 1$ is a parameter to denote the efficiency
of particle acceleration)
and the electron cooling time scale $t'_{c} = 3 m_e c/4\gamma_e
\sigma_{\rm T} U'_B$ (where $\sigma_{\rm T}$ is the Thomson
scattering cross section, and $U'_B = {B'}^2/8\pi$ is the 
comoving magnetic field energy density). 
This gives 
\begin{equation}
\gamma_{e,M} = \left(\frac{6\pi e}{\sigma_{\rm T} \kappa B'}
\right)^{1/2} \simeq 1.2\times 10^8 \kappa^{-1/2} {B'}^{-1/2}~.
\end{equation}
Noticing that the synchrotron spectral function $F(x)$ has 
a maximum value at $x=0.286$ \citep{rybicki79}, the maximum electron
synchrotron emission energy is
\begin{eqnarray}
E_{M} & = & 0.286 \frac{3}{4\pi} \hbar \Gamma \gamma_{e,M}^2 \frac{eB'}
{m_e c} (1+z)^{-1} \nonumber \\
& \simeq & 13~{\rm GeV}~ \Gamma_3 \kappa^{-1} 
\left(\frac{1+z}{5.35}\right)^{-1}~.
\end{eqnarray}
The expression is similar to the shock acceleration case
\citep[e.g.][]{wang09}, although $\kappa$ can vary depending
on the instability growth rate in the ICMART cascade.
In any case, the observed
maximum photon energy 13.2 GeV can be interpreted given a large
enough $\Gamma_3 > 1$ and a not too large $\kappa$. 
A large $\Gamma$ is consistent with the constraint derived from 
the opacity argument \citep{zhangpeer09}.
Alternatively, photons above 100 MeV may be originated from the
external shock \citep{kumar09,ghisellini09}\footnote{This requires
a high degree of coincidence since a detailed time-dependent
spectral analysis with as many time bins as possible still reveals
a series of nearly featureless Band-function spectra for all the
time bins \citep{zhang10}.
In any case, all the arguments for the 
ICMART model discussed in this paper are not affected even if 
$>100$ MeV photons are of the external shock origin.}.
The requirement for $\kappa$ and $\Gamma$ 
is then much less demanding. 

Another interesting feature of GRB 080916C is the delayed onset
of the LAT band emission. In fact, most of the multi-component models
discussed in the literature \citep{toma09,li10,razzaque09} are
motivated to interpret this feature. To us, this feature may
be straightforwardly interpreted in the following way. The particle 
acceleration details could be different during the first ICMART 
event than the later ones. Either
the electron spectral index is steeper, or there is a
pair cutoff feature in the LAT band (the latter may be related
to a smaller $\Gamma$ or a smaller emission radius). 

\section{Conclusions and Discussion}\label{sec:conclusion}

We have developed a GRB prompt emission model in the high-$\sigma$
regime, namely, the Internal-Collision-induced MAgnetic Reconnection and
Turbulence (ICMART) model.
This model is motivated by the Fermi observations of GRB 080916C
\citep{abdo09}, and developed upon the idea that GRBs are powered by 
3D turbulence reconnection discussed by \cite{lazarian03} (but in
the high-$\sigma$ regime).
This model inherits the merits
of the internal shock and other models, but may overcome several 
drawbacks of the internal shock model (low efficiency, fast cooling,
electron number excess, Amati/Yonetoku relation inconsistency, 
missing bright photosphere component, etc). The basic ingredients of 
the model include the following.
\begin{itemize}
\item The outflow launched from the GRB central engine has
a high magnetization, i.e. $\sigma \gg 1$. The $\sigma$ value 
does not decrease significantly before reaching the GRB
emission radius $R_{\rm GRB}$, so that at $R_{\rm GRB}$,
one still has $1 \lesssim \sigma \lesssim 100$. 
As a result, the photosphere 
emission is not bright, and the observed emission is dominated
by the non-thermal emission released during the ICMART events.
\item The central engine is intermittent, launching an unsteady
wind with variable luminosity and Lorentz factor. The wind
interacts internally via collisions.
\item Most collisions at small radii from the central engine
do not result in significant energy dissipation. They mainly
serve to distort the field lines, making the field configuration
progressively irregular.
\item At a certain large radius, 
the condition for turbulent magnetic field reconnection 
is satisfied. Reconnection events rapidly eject energetic particles
to the ambient, which further 
drive turbulence. This results in a run-away discharge of the
magnetic field energy in a reconnection/turbulence avalanche. 
This is one ICMART event, which corresponds to one broad
GRB pulse. During the magnetic field energy discharge, the
$\sigma$ value drops from the original value to around unity.
\item A GRB is composed of several ICMART events (i.e. broad 
pulses), each marks a catastrophic event to destroy ordered magnetic 
field lines. The peak energy $E_p$ is expected to drop from
high to low across each pulse. The $\gamma$-ray polarization degree 
is also expected to drop from $\sim 50-60\%$ to $\sim$ a few
$\%$ during each pulse. The magnetic field configuration at 
the end of prompt emission is largely randomized, but still 
has an ordered component. The reverse shock emission is
expected to be moderately polarized. 
\item The GRB light curves should have two variability components,
a broad (slow) component related to the central engine activity,
and a narrow (fast) component associated with the relativistic
magnetic turbulence.
\end{itemize}

This model differs from other magnetic GRB prompt 
emission models proposed in the past. The EM model proposed
by \cite{lyutikov03} invokes an extremely high-$\sigma$
($\sigma > 10^6$) at the deceleration radius, which might 
not be realized in nature. The 
variability in this model has no direct connection with the 
central engine activity, while the evidence of an 
engine-related variability (e.g. those in X-ray flares) is 
mounting. A large number of GRB magnetic models are in
the MHD regime 
\citep{thompson94,spruit01,drenkhahn02,vlahakis03,giannios08,komissarov09}.
These models invoke magnetic dissipation at smaller radii
to enhance the photosphere emission. At large radii, it is
assumed that the outflow is no longer Poynting-flux-dominated,
so that the internal shock model can still operate. 
On the other hand, in the ICMART model it is envisaged that
rapid reconnection/turbulence cascade only happens under a 
certain trigger condition, preferably at a large emission
radius when the field lines are sufficiently distorted.
So the main difference between the ICMART model and other
MHD models is whether the magnetic energy is released 
abruptly at a large radius or continuously at small radii.

The physics invoked in this model is complicated. In this paper, 
we only limit ourselves to an analytical, qualitative delineation
of the general picture of the model. Many ingredients of the model, 
such as magnetic acceleration in a high-$\sigma$ flow, collision 
physics of high-$\sigma$ shells (shocks and magnetic field distortion),
reconnection physics, particle acceleration (1st-order vs.
2nd-order Fermi acceleration) and radiation, are introduced
based on the best known results from the literature. Many
speculations are subject to quantitative analyses and numerical 
simulations to verify. Further investigations are needed and
indeed in plan.

Within the ICMART model, since the baryon number is smaller
by a factor of $(1+\sigma)^{-1}$ than the pure baryonic model,
the expected hadronic radiation is also smaller by the same
factor. Internal shocks have been proposed as
the source of ultrahigh energy cosmic rays \citep[UHECRs,][]{waxman95}
and PeV neutrinos \citep{waxman97}. The ICMART processes can
in principle also accelerate protons to ultrahigh energies.
However, the UHECR flux from a high-$\sigma$ GRB is smaller by a 
factor of $(1+\sigma)^{-1}$ than a baryon-dominated GRB. 
If the majority of GRBs have high-$\sigma$, then 
GRBs cannot be the dominant contributor to the observed 
UHECRs in the solar neighborhood. Similarly, the predicted
diffuse PeV neutrino background is also lowered by a factor
of $(1+\sigma)^{-1}$. This makes it more challenging for km$^3$ 
neutrino telescopes (e.g. Icecube) to detect these neutrinos 
from GRBs\footnote{Our discussion is relevant to traditional
high luminosity GRBs. The nearby low-luminosity GRBs may
be more abundant \citep{liang07,virgili09}. So far there is no 
strong evidence that they are Poynting flux dominated. If they
are matter-dominated, they can be important contributors to
diffuse high energy neutrinos and UHECRs 
\citep{gupta07,murase06,murase08}.}.

On the other hand, the GRB composition may be diverse, namely,
the $\sigma$ value may vary in a wide range among GRBs.
For example, another LAT burst GRB 090902B shows a clear
time-evolving blackbody component superposed on a non-thermal
power law component \citep{ryde10,zhang10}, which is almost certainly
the baryonic photosphere component. This burst is likely
originated from a baryonic fireball proposed by \cite{paczynski86}
and \cite{goodman86}, see \cite{peer10} 
for a more detailed discussion. However, GRB 090902B is a special 
event. Its special spectral feature is unique within the Fermi
LAT GRB sample of \cite{zhang10}. Most LAT GRBs have clear
Band-only time-resolved spectra similar to GRB 080916C, which are good 
candidates for the ICMART scenario \citep{zhang10}. The ICMART model 
proposed here therefore can be applied to most GRBs. 

The ICMART scenario developed in this paper may be also applied
to other astrophysical objects, such as active galactic nuclei
(AGNs). For example, two blazars (Mrk 501 and PKS 2155-304) were
detected to have a 3-5 minute TeV variability
\citep{albert07,aharonian07}, much shorter than the inferred 
light-crossing times at the black hole horizon. The detections
of the TeV photons also require a Lorentz factor much larger 
than that inferred from the large scale jet modeling. The data
demand some small-scale enhanced emission units 
\citep{begelman08,giannios09b}. Suppose that an intermittent, 
moderately high-$\sigma$ outflow is launched by the supermassive
black hole central engine from these blazars, ICMART
events similar to what are discussed in this paper
may occur, which would produce small-scale turbulent
emission units whose size is much smaller than the black hole
event horizon. These turbulent eddies are relativistic, and 
hence, would give an extra Lorentz boost to the comoving 
emission. This would account for the observed rapid TeV
variability and the apparent large Lorentz factor of the TeV
emission regions of these two blazars.

\begin{acknowledgements}
This work is supported by NASA NNX09AT66G, NNX10AD48G, NNX10AP53G, 
and NSF AST-0908362 (BZ) and by a 985 grant at Peking University
and Arizona Prize Fellowship (HY). 
We thank the anonymous referee for constructive
comments. We also acknowledge helpful communications with or comments 
from the following colleagues on various topics discussed in this paper: 
J. Arons, A. M. Beloborodov, R. D. Blandford, Z.-G. Dai, 
F. Daigne, E. M. de Gouveia Dal Pino, Y.-Z. Fan, Z. Li, E.-W. Liang, 
S. Kobayashi, S. S. Komissarov, A. K\"onigl, P. Kumar, A. Lazarian, 
D. Lazzati, M. Lyutikov, J. C. McKinney, M. V. Medvedev, P. M\'esz\'aros, Y. Mizuno, 
S. Nagataki, E. Nakar, R. Narayan, K.-I. Nishikawa, R. Ouyed, A. Pe'er, 
T. Piran, D. Proga, J. Poutanen, S. Razzaque, A. Spitkovsky, K. Toma, 
X.-Y. Wang, X.-F. Wu, and B.-B. Zhang. 
\end{acknowledgements}


\begin{appendix}
\section{Notation list}
The notation we used is listed in Table 1.

\begin{table}
\begin{tabular}{ll}
$c$ & speed of light \\
$c_s$ & speed of sound \\
$d$ & separation between shells in the lab frame, $=c \delta t$ \\
$d_{\rm max}$, $d_{\rm min}$ & maximum and minimum separation between shells \\
$e$ & electron charge \\
$h$ & Planck constant \\
$\hbar$ & reduced Planck constant \\
$k$ & Boltzmann constant or turbulence wave number \\
$k_\perp$, $k_\parallel$ & MHD turbulence wave number in the direction perpendicular/parallel to the magnetic field \\
$l$, $l'$ & mean free path of microscopic interactions in the lab frame and comoving frame, respectively\\
$l'_{e,col}$ & comoving mean free path of electron Coulomb collision \\
$m$ & a dimensionless parameter defined as $1+Y m_e/m_p$ \\
$m_1$, $m_2$ & mass of shell 1 and shell 2\\
$m_e$, $m_p$ & electron rest mass, proton rest mass \\
$n$ & particle number density of the GRB outflow in the lab frame \\
$n_e$, $n'_e$ & electron number density in the lab frame and comoving frame \\
$n_p$, $n'_p$ & proton number density in the lab frame and comoving frame \\
$n_{_{\rm GJ}}$ & Goldreich-Julian charge number density of the GRB outflow in the lab frame \\
$p$ & power law spectral index of particles (electrons or protons) \\
$r'_{B,e}$, $r'_{B,p}$ & comoving electron gyro-radius and proton gyro-radius \\
$r_{col}$ & strong Coulomb collision radius \\
$s$ & dimensionless parameter defined in Eq.(\ref{tau}) \\
$t'_{acc}$ & comoving particle acceleration time scale \\
$t'_c$ & comoving cooling time scale \\
$t'_{dyn}$ & comoving dynamical time scale \\
$v'_e$ & comoving electron speed \\
$z$ & redshift \\
$B$, $B'$ & magnetic field strength in the lab frame and comoving frame \\
${\bf B}$ & magnetic field vector in the lab frame \\
$B_q$ & critical magnetic field strength \\
$E$ & observed photon energy \\
$E_h$ & the critical photon energy below which slow heating effect is important \\
$E_M$ & maximum observed photon energy \\
${\bf E}$ & electric field vector in the lab frame \\
$E_p$ & observed spectral peak energy of GRB \\
$E_{\gamma,iso}$ & isotropic gamma-ray energy of GRB \\
$E(k)$ & turbulence energy per unit wave number at the wave number $k$ \\
$E(k_\perp)$, $E(k_\parallel)$ & turbulence energy per unit wave number at the wave number $k_\perp$, and $k_\parallel$ \\
$E(\theta)$ & GRB effective ``isotropic" energy at an angle $\theta$ from the jet axis \\
$\dot E'_{heat}$, $\dot E'_{cool}$ & comoving average heating and cooling rate of electrons \\
$F(x)$ & a function to denote synchrotron emission spectrum of a single particle \\
$F_b$ & matter flux: sum of baryonic and leptonic
fluxes. Usually baryonic flux dominated. \\
$F_P$ & Poynting flux \\
${\bf J}$ & current density vector \\
$L$ & characteristic length scale of the flow \\
${\cal L}$ & length of reconnection layer \\
$L_w$ &  the total (including kinetic and magnetic) isotropic luminosity of the GRB ejecta (wind) \\
$L_\gamma$ & the isotropic $\gamma$-ray luminosity of the GRB \\
$N(E)$ & number of photons in the energy bin $(E, E+dE)$ \\
$N(\gamma_e)$ & number of electrons in the Lorentz factor bin $(\gamma_e, \gamma_e+d\gamma_e)$ \\
$P$ & pressure \\
$P_{gas}$ & gas pressure \\
$P_{mag}$ & magnetic field pressure \\
$P_{ram,21}$ & ram pressure exerted to shell 1 by shell 2 \\
$R$ & radius from the central engine \\
$R_{\rm dec}$ & GRB ejecta deceleration radius \\
$R_e$ & Reynold's number \\
$R_{\rm GRB}$ & radius of GRB prompt emission from the central engine \\
$R_{\rm ICMART}$ & radius of the ICMART events \\
$R_{\rm IS}$ & internal shock radius defined in Eq.(\ref{RIS}) \\
$R_m$ & magnetic Reynold's number \\
$R_{_{\rm MHD}}$ & radius from central engine where the MHD condition is broken \\
$R_\theta$ & jet cross section radius at the radius $R$ \\
$S$ & Lundquist number \\
$T$ & characteristic temperature of plasma \\
$T_e$, $T_p$ & characteristic electron temperature, proton temperature \\
$T_{e,c}$ & critical electron temperature for collisional/collisionless regime separation \\
$U'$ & total internal energy in the internal shock or ICMART event \\
$U'_B$ & comoving magnetic field energy density \\
$U'_{ph}$ & comoving photon energy density \\
${\bf V}$ & velocity vector of the GRB fluid \\
$V_{\rm in}$ & incoming speed of the magnetic field lines in Sweet-Parker reconnection \\
$V_{\rm out}$ & the eventual outgoing speed in a reconnection event \\
$V_{\rm A}$ & Alfv\'en speed in general \\
$V'_{\rm A}$ & comoving Alfv\'en speed \\
$V'_{\rm A,NR}$ & comoving Alfv\'en speed in the non-relativistic regime \\
$V'_{\rm rec,loc}, V'_{\rm rec,global}$ & comoving local and global reconnection speed \\
$x$ &  argument of function $F(x)$ \\
\end{tabular}
\end{table}
\begin{table}
\begin{tabular}{ll}
$Y$ & number of leptons associated with each proton \\
${\cal Y}$ & The Compton parameter \\
$\alpha$ & GRB photon spectral index below $E_p$ \\
$\beta$ & dimensionless speed $V/c$ or ratio between gas pressure and magnetic field pressure \\
$\beta_f$ & dimensionless speed of the fast shell \\
$\beta_m$ & dimensionless speed of the merged shell \\
$\beta_s$ & dimensionless speed of the slow shell \\
$\gamma_A$ & Alfv\'en Lorentz factor in general \\
$\gamma'_{\rm A}$ & comoving Alfv\'en Lorentz factor \\
$\gamma_e$ & comoving electron Lorentz factor \\
$\gamma_{e,c}$ & comoving electron Lorentz factor at cooling break \\
$\bar\gamma_e$ & comoving mean electron Lorentz factor \\
$\gamma_m$, $\gamma_M$ &  minimum and maximum Lorentz factor of a power-law distributed proton or electron population \\
$\gamma_{e,m}$ & minimum comoving electron Lorentz factor in an injected power law energy spectrum \\
$\gamma_{e,M}$ & maximum comoving electron Lorentz factor in an injected power law energy spectrum \\
$\gamma_{e,p}$ & comoving electron Lorentz factor that contributes to $E_p$ in the synchrotron radiation model \\
$\gamma_p$ & comoving proton Lorentz factor \\
$\bar \gamma_p$ & comoving mean proton Lorentz factor \\
$\gamma_{in}$ & Lorentz factor of incoming magnetic field lines for relativistic reconnection \\
$\Gamma$ & GRB ejecta bulk Lorentz factor \\
$\Gamma_0$ & the ejecta bulk Lorentz factor at the central engine \\
$\Gamma_1$, $\Gamma_2$ & Lorentz factor of shell 1 and shell 2 \\
$\Gamma_{21}$ & relative Lorentz factor between shell 2 and shell 1 \\
$\Gamma_{43}$ & relative Lorentz factor between regions 4 (unshocked trailing shell) and 3 (shocked trailing shell) \\
$\Gamma_f$ & Lorentz factor of the fast shell \\
$\Gamma_m$ & Lorentz factor of the merged shell \\
$\Gamma_s$ & Lorentz factor of the slow shell \\
$\Gamma_{fs}$ & relative Lorentz factor between the fast and slow shells \\
$\Gamma_{ud}$ & relative Lorentz factor between upstream and downstream \\
$\Gamma_{\rm max}$ & maximum Lorentz factor in the GRB ejecta \\ 
$\Gamma_{\rm min}$ & minimum Lorentz factor in the GRB ejecta \\
$\Gamma_{\rm tot}$ & achievable Lorentz factor of high-$\sigma$ shell for total conversion of Poynting energy to kinetic energy \\
$\delta$ & thickness of reconnection layer \\
$\delta'_e$, $\delta'_p$ & comoving electron and proton plasma skin depth \\
$\delta t$ & duration between the end of ejecting a leading shell and the beginning of ejecting a trailing shell \\
$\delta t_{\rm max}$ & maximum $\delta t$ in the ejecta \\
$\delta t_{\rm min}$ & minimum $\delta t$ in the ejecta \\
$\delta V$ & relative velocity of the flow \\
$\Delta$ & shell width in the lab frame \\
$\Delta'$ & shell width in the comoving frame \\
$\Delta_f$ & width of the fast shell \\
$\Delta_s$ & width of the slow shell \\ 
$\Delta_{\rm max}$ & maximum shell width in the GRB ejecta, $=c \Delta t_{\rm max}$ \\
$\Delta_{\rm min}$ & minimum shell width in the GRB ejecta, $=c \Delta t_{\rm min}$  \\
$\Delta t$ & duration of central engine activity for each mini-shell
in the ejecta \\
$\Delta t'$ & comoving time scale for global magnetic dissipation within a shell with comoving width $\Delta'$ \\
$\Delta t_{\rm max}$ & maximum $\Delta t$ in the ejecta \\
$\Delta t_{\rm min}$ & minimum $\Delta t$ in the ejecta \\
$\epsilon_B$ & fraction of internal energy that is distributed to magnetic fields in internal shocks \\
$\epsilon_e$, $\epsilon_p$ & fraction of internal energy that is distributed to electrons and protons in internal shocks \\
$\varepsilon_e$, $\varepsilon_p$ & fraction of dissipated magnetic energy distributed to electrons and protons in an ICMART event \\
$\eta$ & magnetic diffusion coefficient or energy dissipation efficiency in general \\
$\eta_{_{\rm IS}}$ & energy dissipation efficiency of an internal shock \\
$\eta_{_{\rm ICMART}}$ & energy dissipation efficiency of an ICMART event \\ 
$\theta$ & angle from the GRB jet axis \\
$\theta_j$ & GRB jet opening angle \\
$\kappa$ & a parameter to denote efficiency of particle acceleration \\
$\lambda_B$ & coherence length of a random magnetic field \\
$\lambda_\parallel$ & local reconnection length \\
$\nu$ & kinematic viscosity \\
$\xi$ & fraction of comoving magnetic field energy that is
dissipated through slow heating mechanism \\
$\Pi$ & linear polarization degree \\
$\rho$ & mass density in the lab frame \\
$\rho'$ & mass density in the comoving frame \\
$\rho'_1$, $\rho'_2$ & comoving mass density of the shell 1 and shell 2 \\
$\sigma$ & magnetization parameter as defined in Eq.(\ref{sigma}) \\
$\sigma_0$ & initial $\sigma$ value at the GRB central engine \\
$\sigma_c$ & critical $\sigma$ value to separate the sub-Alfv\'en regime from the super-Alfv\'en regime \\
$\sigma_{end}$ & final $\sigma$ value after an ICMART event \\
$\sigma_{ini}$ & initial $\sigma$ value before an ICMART event \\
$\sigma_T$ & Thomson scattering cross section \\
$\tau'_{col}$ & comoving Coulomb collision time \\
$\tau'_{col,NR}$ & non-relativistic comoving Coulomb collision time \\
$\tau'_{col,R}$ & relativistic comoving Coulomb collision time \\
$\tau_\nu$ & viscous diffusion time \\
$\tau_{dif}$ & magnetic resistive diffusion time \\
$\tau_f$ & flow time scale \\
$\phi(p)$ & a function of $p$ to connect minimum particle energy with the mean energy \\
$\omega'_{B,e}$, $\omega'_{B,p}$ & comoving electron and proton gyro-frequency \\
$\omega'_{p,e}$, $\omega'_{p,p}$ & comoving electron and proton plasma frequency \\
\end{tabular}

\end{table}
\end{appendix}



\newpage
\bibliographystyle{apj}

\bibliography{ms}

\bigskip
Notes added in proof: After acceptance of our paper, we were notified by Jon McKinney 
about an alternative mechanism to trigger fast reconnection at
large radii of GRBs (J. C. McKinney \& D. A. Uzdensky, 
MNRAS, submitted, arXiv:1011.1904).

\end{document}